\documentclass[12pt]{article} 
\usepackage{epsf,fancyheadings,amsfonts}   
\usepackage{floatflt,graphicx}
\def\bea{\begin{eqnarray}}
\def\eea{\end{eqnarray}}
\def\be{\begin{equation}}
\def\ee{\end{equation}}
\def\nn{\nonumber}

\def\a{\alpha}
\def\b{\beta}

\def\d{\delta}

\def\e{\epsilon}

\def\g{\gamma}
\def\G{\Gamma}
\def\k{\kappa}
\def\k{\kappa}
\def\l{\lambda}

\def\m{\mu}
\def\n{\nu}

\def\s{\sigma}
\def\t{\tau}
\def\th{\theta}

\begin{document}
\begin{center}
{\hfill \tiny{Preprint no: ULB-TH/03-08}}\\
{  \hfill \tiny{ UNB Technical Rep: 03-02}}
\vspace{1cm}

{\Large Coherent States for Black Holes}

\vspace{1cm}

Arundhati Dasgupta \footnote{adasgupt@ulb.ac.be, dasgupta@aei.mpg.de},

\vspace{0.5cm}
{\it Physique Theorique et Mathematique,\\Universite Libre de Bruxelles, B-1050, Belgium.\\ and\\
Department of Mathematics and Statistics,\\University of New Brunswick, Fredericton, E3B 5A3, Canada.}
\end{center}

\vspace{1cm}

{\bf Abstract}: We determine coherent states peaked at classical
space-time of the Schwarzschild black hole in the frame-work of canonical quantisation of general relativity. The information about the
horizon is naturally encoded in the phase space variables, and 
 the perturbative quantum fluctuations
around the classical geometry depend on the distance from the horizon. For small black holes, space near the vicinity of the singularity appears discrete with the singular point excluded from the spectrum.

\section{Introduction}
Quantum Black holes are still extremely mysterious, even after a decade of optimism, which arose with the microscopic counting
of Entropy.  What is the horizon in the microscopic picture? What happens to the singularity at the centre of the black hole?
What is exactly Hawking radiation, and where is the end point, in a pure or mixed state? There is an immense amount of literature (\cite{bhs,stcv,can1,can})
trying to address these issues, using various approaches, and reviewing them all is impossible here. But what emerges from them,
is that we need a microscopic theory, which is equipped enough to address non-perturbative gravity, i.e. not a theory of gravitons,
but a theory incorporating all the non-linearity associated with strongly gravitating systems like the black hole, also the quantum
states must have a suitable classical limit so that one can identify continuum geometry. In otherwords, a non-perturbative quantum theory, where one can define $h\rightarrow 0$, without making the gravitational fields weak. String theory as it stands today, does not have a
 non-perturbative, back ground independent description so far. Other approaches to quantisation, involving
Path-integrals and discretization \cite{dis}, may have useful models, but are yet unexplored. This brings us to a very interesting construction, in the framework of canonical quantisation of general relativity, of coherent states, which supposedly in any quantum theory
encode the information of classical objects. Thus, we have a theory of non-perturbative quantum gravity, which is back ground independent, 
and a construction which allows one to look for the classical limit. Hence one should immediately apply them to black holes and investigate the appearance of the horizon and singularity in their framework.
The task is thus two prong, trying to understand what coherent states are in the context of gravity, and their
application to extract information about a classical black hole.

\subsection{Coherent States}
In quantum mechanics, coherent states appear as eigenstates of the annihilation operator, and in these the minimum uncertainty
principle, i.e., $\Delta x \Delta p =  h/2$ is maintained by the position and momentum operators. This definition has a generalisation in terms of a `coherent state transform' defined thus: Given a space $L^2(R^n)$, then there exists a map or an integral transform to the space ${\cal H} (C^n)$ of holomorphic functions on $C^n$ \cite{hall}. In other words:
\be
A(f)(z) = \int_{R^n} A (z, x) f(x) dx 
\ee

Where $ A(z, x) = \frac{1}{(2\pi)^{n/4}} \exp\left[ -\frac14 \Sigma (2 z_i^2 + x_i^2 ) + \Sigma x_i z_i\right]$, in case of $i=1$,
this is same as the simple harmonic oscillator coherent state wave function. The above also has a more convenient representation
\be
A(z,x) = \frac{\rho_1 (z-x)}{\sqrt{\rho_1}}
\ee 

With $\rho_t$ being the heat kernel on $R^n$, and $\rho_t(z)$ is the analytic continuation of $\rho_t$ to $C^n$ and $t$ denotes the width of the distribution. This is also called the Segal-Bargmann transform. This definition was generalised for an arbitrary gauge group in \cite{hall}, with the following:
\be
A_t(g,x) = \frac{\rho_t(x^{-1} g)}{\sqrt{\rho_t(x)}}
\label{hall}
\ee
with $x$ belonging to a compact gauge group $K$, $g$ belonging to a complexification of the gauge group, $G$, and the
$\rho_t(x)$ being the heat kernel on $K$ defined with respect to an appropriate measure. Once the $A_t(g,x)$ are determined, they will serve as coherent states
in the `position representation' with width $\sqrt t$, and there will be a corresponding one in the `conjugate momentum'
representation. It is this definition, for the coherent states which can be applied to gravity,
as it appears as a SU(2) gauge theory in the canonical framework. For further work on Coherent
States see \cite{hall2}.

\subsection{Coherent States for Gravity}
In the canonical framework, space-time is separated as $R\times \Sigma$, where $\Sigma$ is a 3-dimensional manifold,
with a given topology. The two physical quantities relevant for this spatial slicing are the triads which determine 
the induced metric and hence the intrinsic geometry of $\Sigma$, and the extrinsic curvature which determines how
the slices are embedded in the given space-time. The triads have an internal degree of freedom associated with the tangent plane. This $SO(3)$ symmetry on the tangent plane, is isomorphic to SU(2), and one can    
combine the variables to give a connection which transforms in the adjoint of the $SU(2)$ gauge group. The precise definitions are:
\be
A^I_a = \G^I_a + \b K_{ab} e^{b I}, \ \ \ \ \ \b E^a_I E^{b I} = \det q q^{a b}
\label{ashbarim}
\ee

where $a$ denotes the spatial indices $1,2,3$ and $I$ the internal SU(2) index also
taking values $1,2,3$. 
 $q_{ab}$ is the induced metric on the three manifold, the $e^b_I$
are the corresponding triads, $\G^I_a$ is the spin connection associated with them. 
$K_{ab}$ is the pull back of the extrinsic curvature on to the spatial slice. The new variables, the connection
$A^I_a$ and the one form $E_a^I$ constitute what is known as the `Ashtekar-Barbero-Immirzi' variables, with
a one parameter ambiguity present in the definition called the Immirzi parameter $\b$. Details of this can be found
in the reviews \cite{rov,thiem,thiem'} and books \cite{pul}. The Einstein-Hilbert action written down in the above variables includes additional constraints associated with the lapse and the shift, as
 well as the Gauss constraints associated with any Yang-Mill's theory. On the constrained space, the `Phase Space' variables satisfy appropriate Poisson brackets. Since the 
quantisation has to be background independent, and non-perturbative, one takes the help 
of `Wilson Loops' which are path-ordered exponentials of the connection, along one-dimensional curves. 
This leads to the introduction of 
 abstract graphs $\G(v,e,I)$ which are comprised of the disjoint sets of vertices $v$, 
edges $e$, and a incidence relation $I$ between them. e.g. A graph with every element of $e$ is incident with two elements of $v$ is a simple planar graph. The restriction on the graphs used in this formalism is the requirement of piecewise analytic edges.
For details, see \cite{rov,thiem,thiem'}. The holonomies along these edges constitute the configuration space variables.
  The quantum Hilbert space constitutes of Cylindrical functions, $f(h_{e_1}(A), h_{e_2} (A), h_{e_3}(A),...,h_{e_i}(A),..)$ where $h_{e_i}(A)$ is the holonomy of the connection along the $i$ th edge. This brings us to the coherent state transform within the framework of loop quantum gravity \cite{Asht,thiem1,thi2,thiow1,thiow2,thiow3}, 
For each edge $e$, the holonomy $h_e(A)\subset SU(2)$, and hence a complexification of K=SU(2), would lead to the complex group
G defined by Hall. The transform would take a function cylindrical over SU(2), to a function of the complexified holonomy.
The transform will have the kernel, using the generalisation by Hall. There is a subtlety in the definition pointed out by B. Hall, where the actual transform used in gravity is the $C_t$ transform defined in Equation C, of Appendix of (\cite{hall}).  
\be
 C_t(g,h) = \sum d_{\pi} e^{- t \lambda_{\pi}} \chi_{\pi}\left[gh^{-1}\right]
\label{coh}
\ee

Where in the above, $h$ is the element corresponding to the holonomy, $g$, the complexified group element, and the RHS is an expansion
of the Heat Kernel on SU(2) based on Peter-Weyl decomposition. Given an irreducible representation $\pi$ of SU(2) or its complexfication SL(2,C),
$d_{\pi}$ is its dimension, $\lambda_{\pi}$ the eigenvalue of the Laplacian, and $\chi_{\pi}$ its character. Thus, the coherent state
on each edge of the graph by above is:
\be
\psi^t_e(g,h) = C_t(g,h)
\ee
The coherent state for the entire manifold is then
\be
\Psi^t_{\G}(g,h) = \prod_e \psi^t_e(g,h)
\label{coh1}
\ee
Thus given the above, it remains to find the complexified group element $g$ for gravity.
Usually the conjugate variable to $A_a^I$, the densitised triad is used to complete the phase-space, but for requirements of a anomaly free algebra, and gauge covariance,
Thiemann has defined a momentum variable 
\be
P^I = -\frac{1}{a_N}Tr \left[ T^I h_{e} \left( \int h_{\rho}(p) E^a h^{-1}_{\rho}(p) \e_{abc} dS^{bc} \right) h_{e}^{-1}\right]
\label{momv}
\ee

The `conjugate momentum' is associated with each edge `e' and evaluated on a $D-1$ `dual' surface, which comprise a Polyhedronal decomposition of the sphere. It does not depend exclusively on the densitised triad, but also on the
connection through the holonomies $h_e$, and $h_{\rho}$ as defined above. 
One thus needs a dual polyhedronal decomposition of $\Sigma$, made up of dim D-1 surfaces, intersected by the
edges transversely. The holonomies $h_e$ are holonomies evaluated on the edge from the 
starting point of the edge to the point of intersection. The $h_{\rho}$ are holonomies
 along edges confined to the dual surface, extending from a given point $p$ to point of 
intersection.  ($T^I$ is the generator) Now, this 
specially smeared operator, has an appropriate Poisson algebra with the holonomy variable $h_e(A)$.

\be
\left\{ h_e, h_{e'}\right\}=0 \ \ \ \ \left\{h_e, P^I_{e'}\right\} = t \delta_{e e'} T^I h_e \ \ \ \ 
\left\{P^I_e, P^J_{e'}\right\} = t\e^{IJK} P^K_e \delta_{e e'}
\label{alg}
\ee
($t= \k/a_N$).
This algebra can be lifted to commutator brackets, where now $t=l_p^2/a_N$ and has a suitable representation.
 Using this 
   and a suitable complexification scheme, Thiemann determined the 
complexified group element to be 
\be
g= e^{i T^i P_i} h_e(A)
\label{comp}
\ee

Note the presence of an undetermined constant $a_N$ which was introduced to
make the quantity P dimensionless. This quantity as noted in \cite{hano}
has an analogy with the Harmonic oscillator frequency, and is a input
in the theory.

Thus, given a graph $\G$, its piecewise analytic edges, the corresponding dual graph, the $h_e,P_e^I$ constitute 
the graph degrees of freedom. The coherent state is obtained as a function of these parameterized as $h_e = e^{\mu^I T^I}, P= P^I T_I$. 
The knowledge of what the classical values for these are once a graph is embedded in a space-time which is a 
solution to Einstein's Equation, the point $g$ at which the state must be peaked is determined through Equation (\ref{comp}). As defined in Equation (\ref{coh}), the coherent state is completely known. Using the above definition for the coherent state, and it's gauge invariant counterpart
obtained by group averaging method, in a series of papers, these coherent states were demonstrated as
having the following: They were peaked at appropriate classical values, they satisfied the Ehrenfest theorem,
 (expectation values of operators were close to their classical values, and one could obtain a perturbation 
series in $t$);  the states are overcomplete with respect to an appropriate measure \cite{thiem1,thi2,thiow1,
thiow2}. Finally to make the definition complete (\ref{coh1}), one must take Infinite tensor product limit to incorporate 
asymptotically flat space-time(also the case relevant here), where the
Graphs are infinite. This was addressed in \cite{thiow2,thiow3}. So,
modulo the dependence on the Graph, and the choice of corresponding dual surfaces, one can believe
that one has found the coherent states for non-perturbative gravity. For a comprehensive discussion,
and other attempts to relate to semiclassical physics see \cite{thie2,other,other2}.
\subsection{For a Black Hole}
What would thus a coherent state for a given classical space-time mean? The states as mentioned earlier
are defined on edges of a given abstract graph. This graph can also be embedded in the classical spatial slices,
and the distances along the graph measured by the classical metric. The information about the classical
holonomy along the embedded edge is encoded in the complexified group element $g$. 
 The behavior of $\psi_t$ as a function of $h$ in the connection representation gives the quantum fluctuations
around the given classical value $g$. The construction of the coherent state for a black hole would thus involve, choice of a suitable slicing of the black hole, embedding the
chosen graph in the slice, and then evaluation of the classical holonomy and the corresponding momentum
(\ref{momv}) along the edges. 

In this article we use the above outlined framework to determine the coherent states for a Schwarzschild black hole. 
The formalism of course can be extended to other black holes with multiple horizons, non-trivial maximally symmetric space-times
like de-sitter space which has a cosmological horizon. Using a suitable choice of coordinates for the static black hole space-time,
we obtain flat spatial slices, and embed spherically symmetric graphs on the slices. All these exersices
are done to extract maximal information in the simplest possible settings, as apriori the task looks quite a difficult one to handle.
Once the classical holonomies and the momenta have been evaluated along the embedded edges, one can write down the wavefunction for each 
edge. Since the position of the embedded edge can be measured by the classical metric, one can comment on how the quantum fluctuations
behave as a function of distance from the center of the black hole. We find that the wavefunction
depends on the position of the horizon, and for a particular orientation of the radial edge,
becomes a gaussian. This radial edge lies along the $\th_0=\pi/2$ plane , and we call it the `ray of least resistance'. We follow this edge deep into the black hole using finer and finer
dual surfaces, and calculate the corrections to the geometry. The idea in this article is to
evaluate the corrections to the momentum and holonomy operator, and since most other
quantities can be defined in their terms, the quantum fluctuations to other operators can be determined using these. The results we find are interesting, but not dramatic. The momentum operator for the
radial edge has the information about the horizon, as expected as the dual surfaces are $r=const$
and hence lie
transverse to a freely falling particle. Depending on the size of the edge, oscillations set in as soon as one crosses the horizon, with decreasing amplitudes. 
The tidal forces acting on the spherical surface distort it, and oscillations arise due to that. 
The frequency increases, but the amplitude of oscillations are as small as $10^{-9}$. The $P$ are measured with 
respect to a given length scale $a_N$ and this is what determines the `semiclassicality of the wavefunction'.
 The semiclassicality parameter is $t = \frac{l_p^2}{a_N}$ and hence the bigger $a_N$ is with respect to the Planck Length, the more is the wavefunction peaked at the classical value. However as we discuss later on the value of the classical momentum falls sharply once inside the horizon, and is proportional to 
$\frac{r_g^2}{a_N}$. Hence to keep the amplitude of oscillations regulated as one falls inside the black 
hole to `semi-classical values', it has to take values much smaller than the black hole size. Optimizing $a_N$ leads to an interesting observation that for black holes which are slightly bigger than Planck size, the semiclassicality parameter is around $1$, preventing a perturbative expansion around a given classical value. 

For larger black holes,
 everything is smooth across the horizon, as one is in a freely falling reference frame. Note at this stage, one is calling the position of the horizon, as the point at which $r=r_g$, i.e. the classical value. This brings us to the derivation of the corrections to geometry, in the present situation, they look perfectly in control. 
We calculate the corrections in the $P^I$ component, for the radial edge, which lies along the path of least resistance. For large $P$, the corrections go as $t(P + 1 + 1/P)$, and for small $P$ as $ t(P^2 + P + 1)$. Since $P$ is well behaved, the corrections are not unbounded anywhere in the configuration representation.

To examine the singularity one has to look into the details of the embedded graph, which taking into account the singularity
of the connection, cannot be embedded classically at r=0, but starts some distance away. The dual graph, however includes the
$r=0$ point, as the densitised triad by itself is not singular at $r=0$ and there is no 
harm in including the same. Thus despite the fact that one starts in configuration space without including the singularity, the dual graph
when embedded has the knowledge about the singular point. Thus, in the vicinity of the singularity it is 
correct to work in the Electric field representation, and that is what we resort to. 

After examining the semiclassical expansion around a given classical value, one can extend the argument to include values of $h,P$ whose value is $\sim t$, (Note $t$ is a dimensionless number).
To do that, and also examine the space near the vicinity of the singularity, one goes to the electric field or momentum representation, and finds that the peak for
small black holes, is restricted to take certain quantised value of the momenta. (Note that it is the dual graph, which can include the singular point, as it is does not diverge there)
For large values of $P$, and large enough black holes, the classical continuum geometry is recovered. However, for smaller black holes, as one approaches the singularity, the momentum is
peaked only at certain discrete values given by $(j + 1/2)t$. The important point is the presence of a non-zero 'remnant' which prevents the momentum operator to be ever peaked at the singular point. So in conclusion, quantisation in terms of the above 
operators, should not see the singularity. One could question the legitimacy of such a strong claim given that one is working
with the simplest graph, but this is related to a property of every single radial edge, which can be part of any graph.

In the next section we begin by slicing the Schwarzschild space-time, then determining the appropriate classical variables. 
In the 3rd section, we work with the simplest possible graph for this spatial slicing, and embed it. After evaluating the classical 
group element, we investigate a single radial edge, which is reminiscent of the straight edge use in loop quantum cosmology \cite{boj,bojo,bojo2}. There the justification used is the isotropy of space, and here, one uses the inherent spherical symmetry. In the next section, the Coherent
state obtained using this edge is investigated in details. In the fifth section we include the angular edges of the graph, and briefly comment on the coherent state for the entire graph. 
Section six involves a conclusion and points for further discussion.

\section{The Classical Preliminaries}
The coherent states as defined are on a spatial slice at a fixed time. 
 So, here we take the static Schwarzschild metric, and
look for a suitable slicing, such that the spatial slices are flat, and time has a definite direction. 
The usual Schwarzschild metric in static coordinates has the following form:
\be
ds^2 = -\left(1 - \frac{r_g}{r}\right)dt^2 + \frac{dr^2}{\left( 1 - \frac{r_g}{r}\right)} + r^2 d\Omega^2
\ee

with time $t$, radial coordinate $r$. It is well known, that this coordinate chart covers only $r> 2GM(=r_g)$.  
And would not be a very convenient one to work with, neither are the $t=const$ slices flat. The other very popular coordinates
are the Kruskal coordinates and then the Eddington-Finkelstein coordinates, both of which are suitable for studying 
null slices.  
The choice of slicing for addressing this problem should be compatible with the canonical frame work, where time is distinguished,
and hopefully cover the entire Black hole space-time. One obvious choice is then the following metric,
where space-time is foliated by a set of proper-time observers hurtling down through the black hole
horizon into the singularity (This set of coordinates are also called the Lemaitre coordinates). 

The metric written in proper-time coordinates is:
\be
ds^2 = - d\t^2 + \frac{d R^2}{\left[\frac{3}{2r_g}(R - \t)\right]^{2/3}} 
 + \left[\frac32 ( R - \t)\right]^{4/3} r_g^{2/3} \left( d\th^2 + \sin^2\th d \phi^2\right)
\ee

with $\t$ denoting the proper-time, and R the spatial direction. If one looks at the coordinate transformations, one finds the following relations
\cite{landau},
\bea 
\sqrt{\frac{ r}{2GM}} dr = \left( dR - d \t\right)  \label{rad}\\
dt = d\tau - f dR , \ \ \ \ \ \ f = \left(\frac{2r_g}{3(R-\t)}\right)^{2/3}
\eea
From (\ref{rad}), it is clear that $r=const$ surfaces are at 45 degrees in the $R-\t$ plane, and the coordinates are smooth across
$r=2GM$.  Also, the singularity of the Black hole is at $R=\tau$. The constant $\tau$ slices are flat,
The slices have no
intrinsic curvature, and this becomes evident by making the transformation (at constant $\tau = \tau_c$)
\be
\frac{dr}{dR} = \frac1{\left[\frac{3}{2r_g}(R - \t_c)\right]^{1/3}}, \label{coord}
\ee
the metric is flat $dr^2 + r^2 d\Omega^2$. The extrinsic curvatures given by
$K_{ij} = -\frac12\partial_\t g_{ij}$ are
\be
K_{RR} = \frac1{2r_g}\frac1{\left[\frac{3}{2r_g} (R - \t)\right]^{5/3}}, K_{\th\th}= - \left[\frac{3}{2r_g}(R - \t)\right]^{1/3}r_g, K{\phi\phi}= K_{\th\th} \sin^2\th.
\label{ext1}
\ee
As expected the extrinsic curvature is singular at the centre of the black hole, showing that the slicing becomes more and more curved in the interior of the horizon.
However, there is apriori no reason to treat the horizon as a boundary as the metric is perfectly smooth across it. However one must add a note of caution
to the above, as the coordinates at present are not complete, in particular, for geodesics starting very far in the past, they only graze the
horizon, never crossing it \cite{fronov}. For the article in hand, this is not a major problem, as we are not addressing global issues like the presence
of `event horizon'. The constant $\t$ slices across some finite range in time are enough for us to build the coherent states, and the information about the
horizon is in the form of the apparent horizon, or the outermost trapped surface. 
The pullback of the above to the spatial slice is very simple to derive from the above:
\be
K_{rr}= \frac12 \left(\frac{\sqrt r_g}{r^{3/2}}\right), K_{\th \th} = - \sqrt{r r_g}, K_{\phi \phi} = -\sqrt{r r_g} \sin^2\th 
\label{ext2}
\ee
Thus, at a constant proper-time $\t_c$, the classical spatial slice is flat as stated earlier, and one proceeds to a determination of the coherent state wavefunction on a given slice $\Sigma_{\t_c}$. Throughout our work, our only input is the classical metric, and we expect the formalism to have answers regarding the geometry and the quantum corrections. Thus, to use the formalism developed in \cite{thiem1,thiow1}, one needs a graph to be embedded in the given spatial-slice, and then evaluation of the $h_e,P^I_e$ with the classical metric in the embedded graph. The coherent states will be peaked at these classical values,
and it is the quantum corrections and the nature of the wavefunction as a function of $h_e,P^I_e$, which crucially depend on the classical metric,
will give us the answer to the question we are asking: Does the quantum geometry know about the horizon, the singularity, existence of trapped surfaces
etc etc. Thus as a begining, one needs to determine what the canonical variables are for the above slicing. As given in Equation (\ref{ashbarim}), one can determine the connection and the densitised triad by using Equation(\ref{ext2}), and the fact that the induced metric is flat.
There is a gauge freedom in choosing the triads, and one can choose them to be diagonal in the indices, thus
\be
e^r_1 = 1 \ \ \ \ \  e^{\th}_2 = \frac{1}{r} \ \ \ \ \ \ \ \ \ \ \ e^{\phi}_3 = \frac{1}{r \sin\th}
\ee

Using the above and (\ref{ext2}), one gets
\be
A^1_r = -\b K_{rr} e^{r1} = - \frac{\b}{2}\left(\frac{\sqrt r_g}{r^{3/2}}\right), \ \ \ A^2_{\th} = \b \sqrt{\frac{r_g}{r}} , \ \ \ A^3_{\phi} = \b \sqrt{\frac{r_g}{r}} \sin\th
\ee
The one forms or the electric fields are:
\be
E^r_1 = \frac{ r^2 \sin \th}{\sqrt \b}, \ \ \ E^{\th}_2 = \frac{r \sin\th}{\sqrt \b}, \ \ \ \ E^{\phi}_3 = \frac{r}{\sqrt \b}
\ee

Thus once the classical variables have been determined, one can proceed to the abstract graph space, which when embedded will
give us the information about the holonomy and the dual momentum. Note that there is additional
contribution to the gauge connection from the spin-connection from the spherically symmetric sector of the metric, which is independent of $r_g$.
\section{The Simplest Graph}
This brings us to the question of what graph we must employ here, so that one can embed it in the given spatial slice.
Since the metric has spherical symmetry, one would presume the use of a spherically symmetric graph. This is motivated from the work on loop quantum cosmology where the reduced Hamiltonian constraint forces the choice of a straight edge \cite{boj,bojo,bojo2}. What would a spherically symmetric graph look like?
Edges which start out from some central point like spokes in a wheel, and
vertices along fixed distances along the spoke. The vertices will constitute a cycle, 
be inter connected by angular edges such that the last edge ends on the first vertex.
 Given the fact that apriori nothing is clear
about what the coherent state is going to look like for the Schwarzschild black hole, one can start with this simple graph indeed.
Enlarging the graph to include more and more edges is not a problem as one can consider one to be a subgraph of the other.
Also, as defined in (\ref{coh1}), the coherent state is obtained as a product of the coherent states for each edge. Hence introducing
more edges contributes to the product. So as a begining, one starts with a 
  planar graph, a extended version of the more well known `Wheel graph'. Which has n 3-valent vertices
with the $i $ th vertex adjacent to the $i-j$ th vertex and the $i+j$ th vertex in the list , and one special vertex also called the hub which is adjacent to all the other n vertices. This wheel graph can be extended to include more
vertices which are adjacent to the previous $n$ vertices constituting a cycle(The initial vertex is same as the terminal vertex). The $i$th vertex in this new cycle is adjacent to the $i -j$ th vertex and $i+j$ th vertex, and to the $ith$ vertex of the previous cycle\footnote{For a introduction to Graph theory
\cite{mathworld}}. 
One can go on adding new cycles. Now, one can add an infinite number of cycles ending up with a countably infinite graph. However, since the embedding will involve the hub
situated at the origin of the singularity, one excludes the hub from the configuration space graph. 

\begin{figure}[t,h]
\centerline{ \epsfxsize 4in
\epsfbox{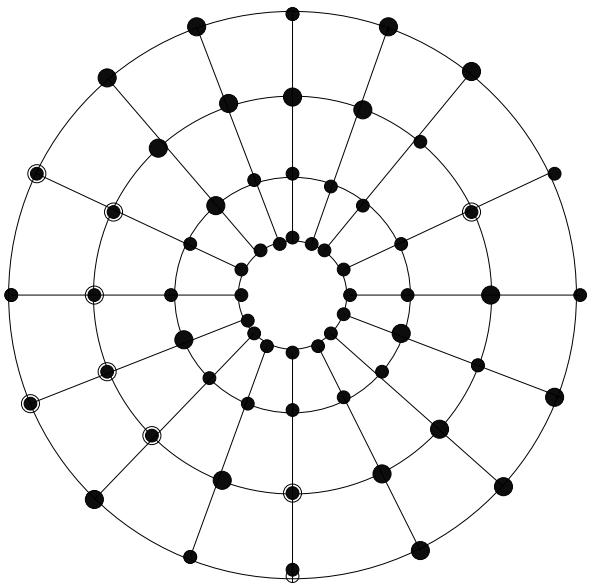}
}
\caption{\sl Spherical Graph}

\label{fig:spg}
\end{figure}
\begin{figure}[t,h]
\centerline{ \epsfxsize 4in
\epsfbox{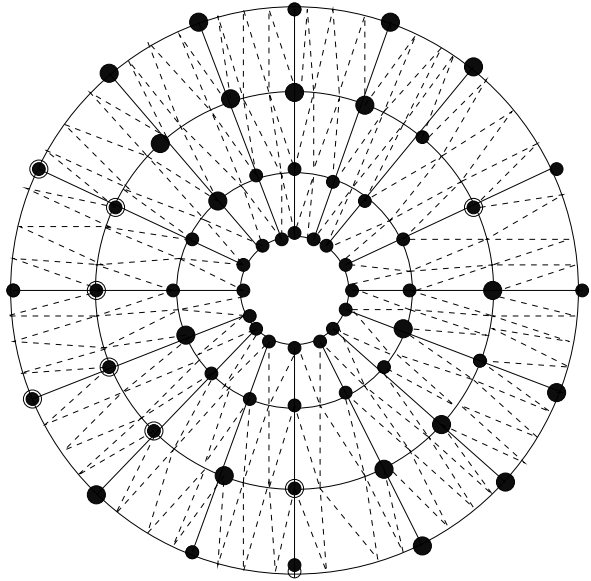}
}
\caption{\sl Non-radial edges}

\label{fig:spgm}
\end{figure}

There are discussions on the exact nature of graph which shall sample maximum number 
of continuum points, \cite{bom}, e.g, one can modify the graph in Figure(\ref{fig:spg}).
Here one has added some more vertices to the cycles, but the edges connecting them
are no-longer be the `spokes of a wheel'. This is as suggested in Figure (\ref{fig:spgm}).

However for our purposes we confine ourselves to the Wheel graph, as this is relevant to us. The embedding of these planar graphs in the $\th=c$ and the $\phi =c$ planes
can give the entire sphere, just like a latitude-longitude grid. A polyhedronal decomposition of the sphere csn be obtained by determining the planar dual and evolving them along the $\phi$ or $\th$ directions. This fixes the `dual' surfaces uniquely. 
The other way of understanding this is that the  planar dual is the intersection of the dual polyhedrons with the corresponding planes in which the planar graphs are embedded. The planar
dual graph corresponding to the graph described earlier is again a set of adjacent cycles, and the hub however reappears here.

\begin{figure}[t,h]
\centerline{ \epsfxsize 4in
\epsfbox{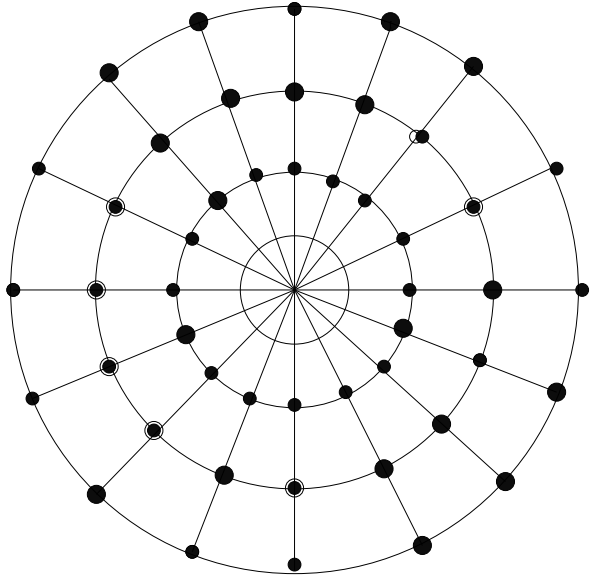}
}
\caption{\sl Dual Spherical Graph}

\label{fig:dspg}
\end{figure}

Question is now, how does one embed this radial spoke into the above spatial-slice. One point to note is that
the $r=constant$ slices for the Schwarzschild radial coordinate, are actually $R-\t$ slices of the Lemaitre metric.
When one is emedding the radial spoke, in the above spatial slice, which is a $\t=constant$ slice, 
and measuring distances in $d r$, defined in (\ref{coord}), one has to keep $\tau_c$ constant, i.e. stay in the same slice. Once the graph is embedded, in the slice, one can determine the holonomy and the momentum (\ref{momv}).
As per the definition (\ref{coh}), one chooses to define the complexified group element $g$ on the embedding of the
edge in the classical space time. This would involve the holonomy and evaluation of the momentum (\ref{momv}) on the
dual surface. 
\subsection{The Complexified group element}
The complexified group element $g$ as stated in Equation (\ref{comp}) has information about the classical holonomy and the corresponding
momentum $P$ evaluated along the dual surfaces.
We thus begin by evaluating the holonomy on the embedded radial edge.

 \subsubsection{The Radial Edge}

\begin{figure}[t,h]
\centerline{ \epsfxsize 4in
\epsfbox{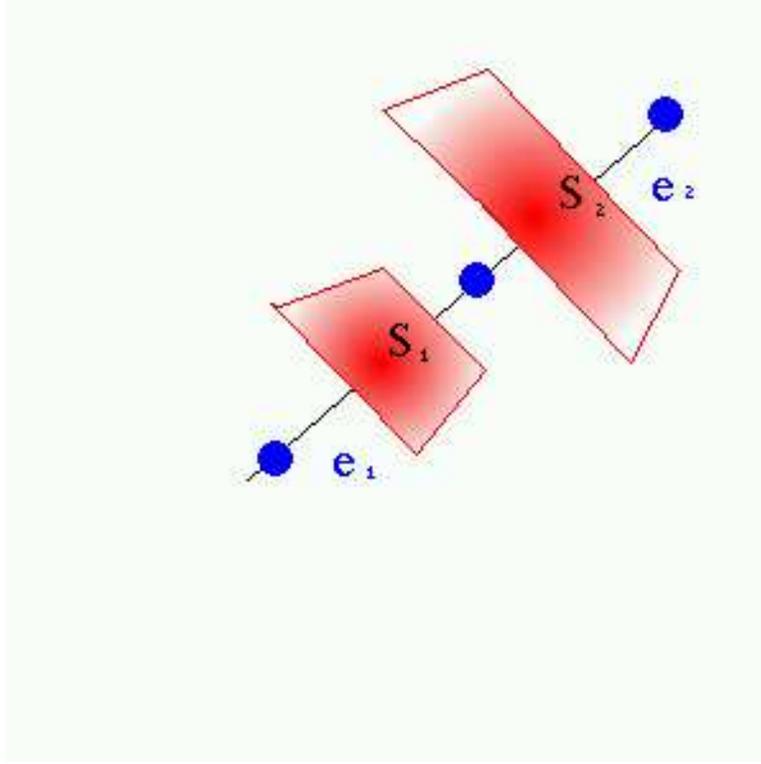}}

\caption{\sl Radial Edge and Dual Surface}

\label{fig:rspg}
\end{figure}

By definition, the holonomy along a given curve (here being the radial edge) is the path ordered exponential of the connection
integrated along the curve. In other words:

\bea
h_e(A)& =& {\cal P}\exp(\int A_{\mu} dx^{\mu})\\
& =& 1 + \sum_{i}\int_a^b d\lambda_1\int_{\l_1}^{b} d\lambda_2...\int_{\lambda_{i-1}}^{b} d\lambda_i A_{\mu}\frac{dx^{\mu}}{d\lambda_i}A_{\mu}\frac{dx^{\mu}}{d\lambda_{i -1}}....
\eea
Where $\lambda$ is the parameter along the curve.
For a radial edge, embedded in the spatial slice, ($\lambda$ is identified with $r$), and due to the
diagonal connection, the holonomy is: 
\bea
h^r_e(A) &=& 1 + \frac{(-\imath \sigma^1)}{2}\int^{r_2}_{r_1} A_r dr + ....\\
&=& \cos\left(\tau\left\{\frac1{r_2^{1/2}} - \frac1{r_1^{1/2}}\right\}\right) - \imath\sigma^1 \sin\left(\tau\left\{\frac{1}{r_2^{1/2}} - \frac1{r_1^{1/2}}\right\}\right)
\label{holr}
\eea
Where $\tau = \frac{\beta}{2} \sqrt r_g$ and the SU(2) generator $T^i= -\frac{\imath}{2} \sigma^i$. The radial edge begins at the point
$r_1$ and ends on the point $r_2$ as measured by the metric on the sphere.
Since only one component of the connection contributes, the subsequent terms commute, and one is left with the above summable series.
Of the few things to observe in the above holonomy calculation is that 
 due to the singularity of the integrand, one in principle must integrate from a region $\e$ distance away from $r=0$. 
So the classical curvature singularity is manifest in the classical holonomy, and whether it is anyway different in
the quantum expectation value is the thing to explore. Next, one must evaluate the corresponding conjugate momentum.
This as defined in Equation(\ref{momv}), involves the choice of a dual surface, and then a choice of edges on the dual surface
which one must integrate over to get the momentum. By definition, this dual surface has to be chosen such that the 
radial edge intersects the surface transversely at a point $p$, only once. At the point $p$, the orientation of the dual surface is in the direction of the edge $e$. All these conditions are satisfied by  pieces of 
spherical surfaces centred at the origin of the radial line. The dual polyhedronal decomposition is thus made 
up of spheres, but now cutting the radial edges at the midpoint. This ensures that the faces of the Dual graph are intersected transversely by the
edges, at precisely one point. The intersection of the dual 2-spheres with the plane of the graph gives the edges of the dual planar graph Figure(\ref{fig:dspg}). The Radial edges  and their duals can thus be visualised as in the Figure (\ref{fig:rspg}).

The $h_{\rho}(\th,\phi)$ are holonomies along a edge connecting a
generic point $p$ on this surface to the point of intersection.  
This choice of the edges on the dual sphere is again difficult to make on the spherical surface, without intr
oducing any ambiguity (see Appendix 2). In the following, the edges are $\th, \phi$ lines, and the shortest distance along the graph edges as an attempt to fix the distances uniquely does not give a meaningful answer as given in Appendix 2. In principle any generic curve connecting the point $(\th,\phi)$ with the point of intersection $p$ would also suffice. In this article, we confine ourselves to the $\phi=0$ plane, and the dual surface, which has to be comprised
of the 2-dim surface is a tiny strip extending from $\phi=0 + \e/2$ to $\phi=0-\e/2$ ($\e$ being a very small number). The intersections of these dual surfaces are 
one-dimensional edges on the $\phi=0$ plane, and these edges comprise the dual planar graph. 
A discussion on the full three dimensional momentum is given in Appendix 1, and this  preliminary simplified 2-surface is sufficient to illustrate
the emergence of the horizon in the variables, and then in the coherent state. The simplified 2-surface has edges along 
the curve $\theta= a \lambda$. Where $\lambda$ is the parameter along the curve and $a$ some undetermined constant. One can think of this as the equator, and the radial edge intersects this
surface once at some point $\theta_0$. The holonomy from any generic point $(\th,0)$ is evaluated along this curve. Eventually one will integrate over the contributions to the holonomy. What we are actually doing in this article is suppressing the azimuthal angle $\phi$, which as we show in Appendix 1 does not change the results drastically. The information that the surface is actually a two sphere is encoded in the
connection, and the one form $E$, which is a densitised triad and includes $\sqrt{q}$, where $q$ is the three metric.

The holonomy on the very thin strip is thus:
\bea
h(\th,\th_0) &= &1 + \int_{\th}^{\th_0} A_{\theta} d\theta + .... \label{holt}\\
&=& \cos\left[\frac{\sqrt{\alpha^2 +1}}{2\sqrt2}\left(\th_0- \th\right)\right] - \frac{\imath}{\sqrt{\a^2 + 1}}(\a\sigma^2 + \sigma^3)\sin\left[\frac{{\sqrt{\alpha^2 +1}}}{2\sqrt 2}\left(\th_0 - \th\right)\right] \nn
\eea
with $\alpha= a\b\sqrt{\frac{r_g}{r}}$. Note now one has taken the contribution from the spin-connection of the spherically symmetric sector into account. We now station the surface at a radial coordinate $r$, and the edge
intersects the surface at the halfway point. The length of the edge is $\delta$. This gives the radial holonomy from Equation (\ref{holr}) with the initial point $r_1 = r-\delta/2$ and the final point $r_2 = r +\delta/2$ for $r\gg \delta$ to be
$h_e(r) = \cos\left(\frac{\tau\delta}{2 r}\right) - \imath \s^1 \sin\left(\frac{\tau\delta}{2r}\right)$. 
Using the above and (\ref{holt}) in (\ref{momv}), and one gets the following expression for the momentum variable :
\be
P^I_r = \frac{1}{a_N}Tr\left[\sigma^{I}\left(\cos\left(\frac{\tau\delta}{2r}\right) - \imath \sigma^1\sin\left(\frac{\tau\delta}{2r}\right)\right)X(r)\left(\cos\left(\frac{\tau\delta}{2r}\right) +  \imath\sigma^1\sin\left(\frac{\tau\delta}{2 r}\right)\right)\right] \label{momv2}
\ee

Where
\bea
X(r) &=& \int_S h (\th,\phi)~ * E ~ h^{-1}(\th,\phi) \nn \\
&= &\int^{\th_0 + \th'}_{\th_0-\th'} h(\th) \frac{r^2\sin\th\sigma^1}{\sqrt{\b}} \nn \\&\times& h(\th)^{-1} d\th 
\eea
(The $\phi$ gives a overall factor of $\e$ which is not shown)
This when evaluated gives
\be
X(r)= X_1(r)\sigma^1 + X_3(r) \frac{1}{\sqrt{\a^2 + 1}}\sigma^2  + X_3 \frac\a{\sqrt{\a^2 +1}}\sigma^3
\ee
with
\bea
X_1(r)&=& \frac{r_g^2}{\a^4} \sin\th_0 \left[ \frac{\sin(1-\alpha')\th'}{1-\alpha'} - \frac{\sin( 1 + \alpha')\th'}{ (1 + \alpha')}\right] \label{int}\\
X_3(r)&=& \frac{r_g^2}{\a^4}\cos\th_0 \left[ \frac{\sin(1-\a')\th'}{(1 - \a')} + \frac{\sin(1+ \a')\th'}{(1 + \a')}\right]  \label{int2}
\eea
Where $\a'= \sqrt{\frac{\a^2 +1}2}$.
Now, one uses the above in (\ref{momv2}), to get the following result for the momentum components
\bea
P^1_r = \frac{ X_1(r)}{a_N} &&  P^2_r =   \frac{X_3}{a_N\sqrt{\a^2 +1}} \left[\a\sin\left(\g\a^3\right) +\cos\left(\g\a^3\right)\right]  \\ &&  P^3_r =  \frac{X_3}{a_N\sqrt{\a^2+1}}\left[- \a\cos \left(\g\a^3\right) + \sin\left(\g\a^3\right)\right] \nn
\label{momb}
\eea

The most important thing to notice in the above is the appearance of the factor $1/(1-\a^2)$ in the above. This for $\b=1/(a)$ denotes the Schwarzschild factor, ($a=1$ and for our purposes we set $\b=1$, it can be restored in the final results, and it's value fixed when comparing to known semiclassical quantities like the
entropy etc)
and when the edge is at $r=r_g$, this factor diverges, signifying a abnormality in the behavior of the function $P^I$. However things are smooth when one takes the
limit carefully due to the presence of $\sin(1-\a')\th'$ in the numerator, which also goes to zero at that point. (We subsequently replace the $\th'\equiv \th,$ and $\delta/2r_g \equiv \g$) 
Note, we did not start working in a
coordinate chart where the horizon was manifest, and it is a nice feature of the above construction that the horizon starts playing some special role. 
Before going into the discussion of what all these quantities mean, one important point to note is the `graph' dependence of the quantities.
The momentum crucially depends on the point at which the radial edge intersects the dual surface, and also the size of the edge, and the dual
surfaces. Thus apart from the dependence on the classical metric, and hence the radial coordinate $r$ at which the edge intersects the dual surface, the size is encoded in $\delta$ and the width of the dual surface in $\th$. In some sense these are 
the graph degrees of freedom, and these additional parameters label the graph, and its particulars. Before going into the construction of the coherent
state, one must have a understanding of the quantity of $P^I$ for the given classical metric, and its behavior.
To do that, we plot the following graphs, illustrating the individual components (in group space) and the gauge invariant quantity $P= \sqrt{P_1^2 + P_2^2
+ P_3^2}$.
Also it is interesting to compare the quantity for the same graph embedded in flat space, with the above as one does not have a natural understanding of the momentum $P^I_e$. 
Thus the momentum variable (for the radial edge) will be
\be
P^1_{\rm flat} \sim \frac{r^2}{a_N} 
\label{momf}
\ee

One sees a distinct difference between (\ref{momb}), and (\ref{momf}), as the first one has `non-abelian' nature, despite the same
graph being embedded, and distinct new length scale in the form of $r_g$. 
 As one can see from (\ref{momf}), this is clearly something which goes as $r^2$, whereas for the black hole,
there is a modulating factor present which we investigate now. For the black hole, where what we call the
$P_{\rm flat}$ ($r^2$) dominates for $0<\a<1$ or outside the black hole, and then around $\a=1$ the modulating factor $P^I/r^2$ start playing a
greater role, taking non-trivial oscillatory nature, with decreasing amplitude. Also, this behavior is not independent of the size of the graph, with the oscillations setting in later and later behind the horizon as the graph is made finer and finer.
So, in some sense $\a=1$ is a transition region in these set of variables, and as we show in the later sections, it is the
manifestation of the momentum variable $P$ in the coherent wavefunction that makes the above analyses worthwhile. Let us start by examining the
gauge invariant quantity $P$. 
\begin{figure}[t,h]
\centerline{ \epsfxsize 4in
\epsfbox{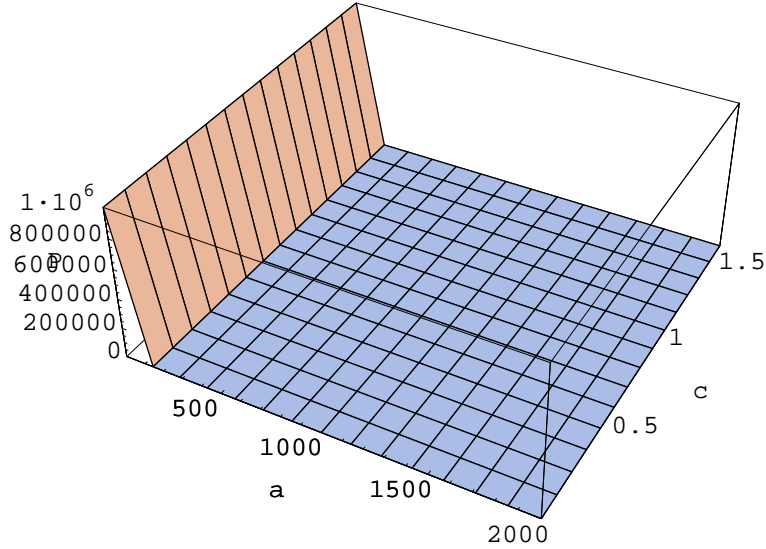}
}
\caption{\sl Gauge Invariant momentum}

\label{fig:gamom}
\end{figure}
\bea
P_{r} &= & \frac{1}{a_N}\sqrt{X_1^2 + X_3^2} \label{momq}\\
&=& 
\frac{r_g^2}{a_N~\a^4}\left[ \left(\frac{\sin[(1-\a')\th]}{(1-\a')}\right)^2 + \left(\frac{\sin[(1+\a')\th]}{(1+\a')}\right)^2 \right.\nn\\ &+& \left.2\cos(2\th_0)\frac{\sin[(1-\a')\th]}{1-\a'}\frac{\sin[(1+\a')\th]}{1+\a'}\right]^{1/2} \nn
\eea

The above is plotted in Figure (\ref{fig:gamom}) (In this and all subsequent Mathematica generated figures,
the angles $\th,\th_0$ appear as $c,c'$, the variable $\a$ as $a$ and the momenta as $P's$ the main reason being Mathematica or mine inability to get greek symbols and superscripts in the exported .eps files).

\begin{figure}[t,h]
\centerline{ \epsfxsize 3in
\epsfbox{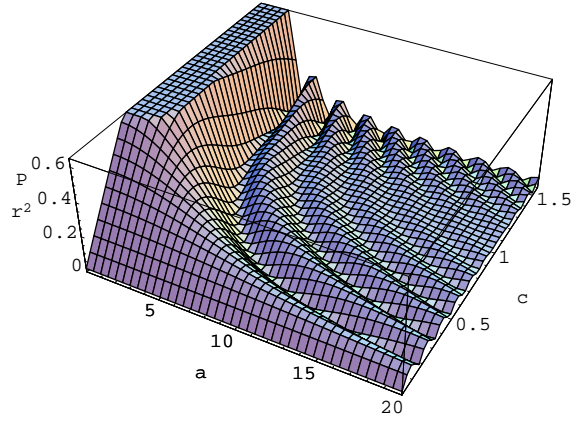}
}
\caption{\sl $P_r/P_{flat}$}

\label{fig:flat}
\end{figure}

(One sets $r^2_g/a_N=200$ as a test and $\th_0=\pi/6$ in the above.)
What is evident from the above is that the fall off of the function as $1/\a^4$ dominates for $0<\a<1$, which is same as the behavior of the
function as would be expected in flat space-time. Just to show, if one divided out the {\it flat} behavior, one would get a modulated graph
as shown in Figure(\ref{fig:flat}).
\begin{figure}[t,h]
\centerline{ \epsfxsize 3in
\epsfbox{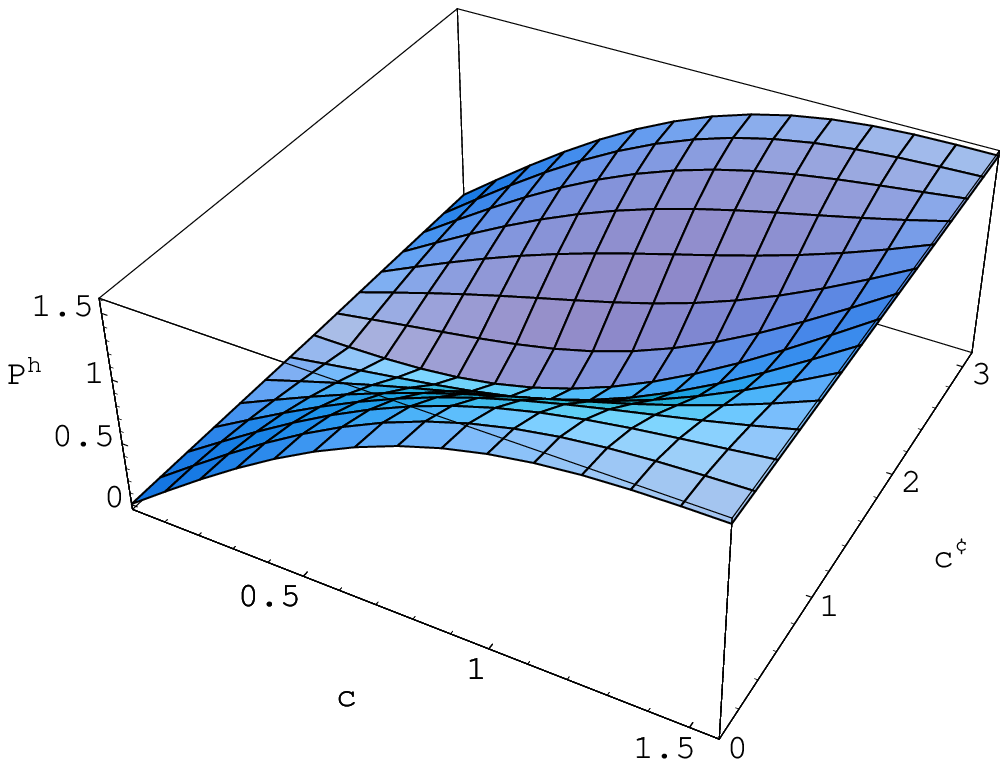}
}
\caption{\sl Momentum at $r=r_g$}

\label{fig:hor}
\end{figure}
Around $\a=1$, one has to take the limit carefully, and one gets the $P_r$ as a function of the angle at which the edge intersects
the dual surface ($\th_0$), as well as the size of the dual surface $\th$. Now, one must notice, that there is nothing drastic in the behavior at the point $\a=1$. This is as plotted as a function of the angle  $\th,\th_0$ (c,c' in Figure (\ref{fig:hor})) at which the radial edge intersects the horizon in Figure(\ref{fig:hor}). 

\begin{figure}[t,b]
\centerline{ \epsfxsize 4in
\epsfbox{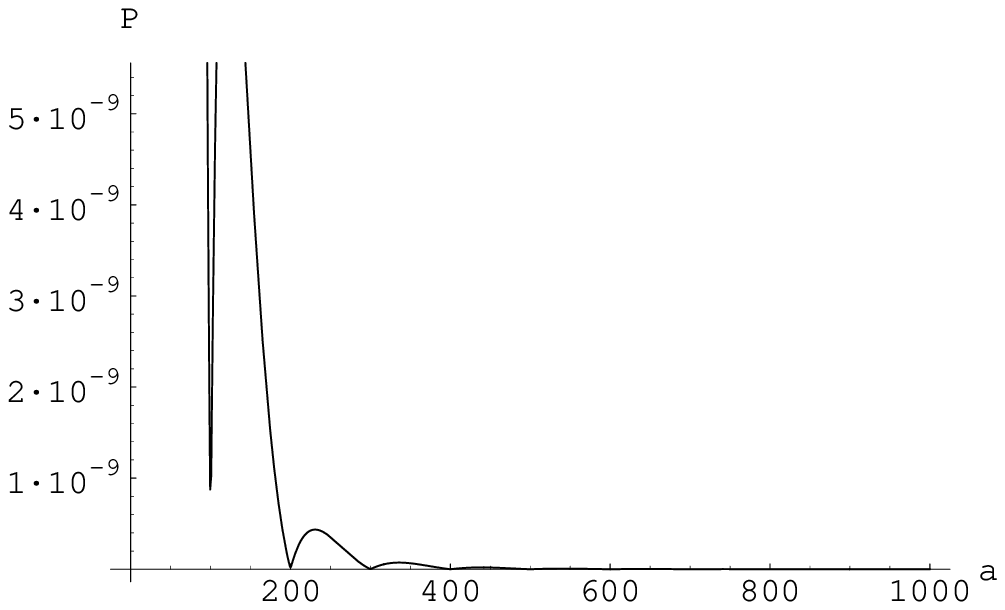}
}
\caption{\sl Finer Graph}

\label{fig:mom34}
\end{figure}

However, around $\alpha >1$ oscillations set in, depending on the size of the dual surface. If the dual surface is big enough (e.g $\th=\pi/2$)
oscillations set in as near as $\a=3$, for finer and finer graphs, $\th=\pi/100$, oscillations set in around $\a=200$. The reason for
this is that in the argument of the $\sin[(1-\a')\pi/N]$ functions, there are two competing factors $1-\a'$ and $\pi/N$. As soon as $1-\a' >N$, 
the oscillations set in. This is an indication of the inability of the surface to be spherical in the increasingly curved 
slice of the black hole. The geodesic deviations become larger and larger, creating the above situation. Below is the plot
for $\th=\pi/100$ Figure (\ref{fig:mom34}).

\vspace{0.5cm}

Similarly, one can obtain the plots of individual components as shown one by one below. 

\begin{figure}
\begin{center}
\parbox{2in}{\epsfxsize 2in 
\epsfbox{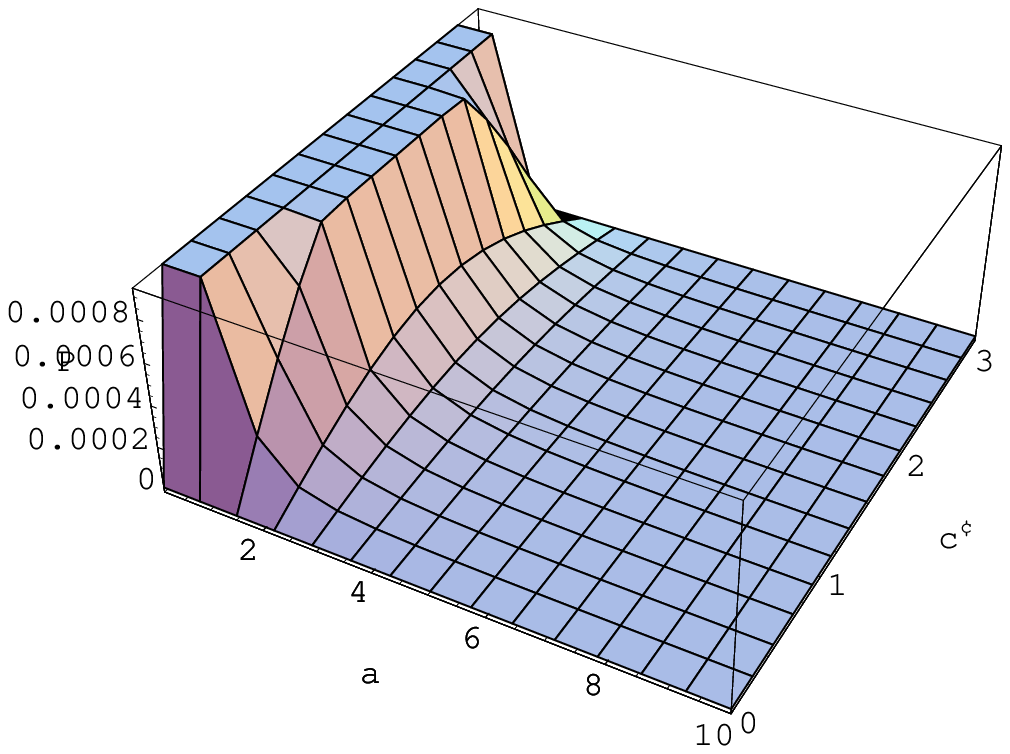}}
\hspace{.25in}
\parbox{2in}{\epsfxsize 2in \epsfbox{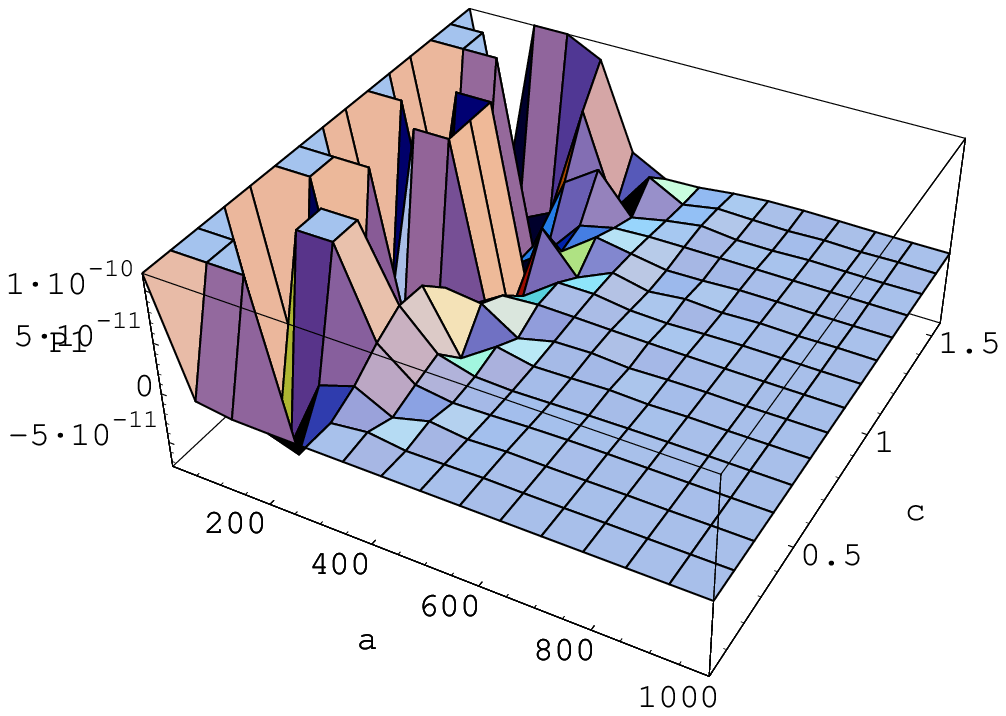}}
\end{center}
\caption{\sl $P_1$: Across the Horizon and Inside the horizon}
\label{fig:mom1}
\end{figure}

This first figure shows the $P^1_r$ component as a plot of $0<\a <10$ denoted as the $a$ axis and $\th$ denotes as the $c$ axis. The value decreases progressively about $\a=1$, it has very low values indeed Figure(\ref{fig:mom1}. 1). 

This next graph shows the same momentum but with $\a$ now in the range (100,1000) Figure(\ref{fig:mom1}.2). The oscillations set in with decreasing amplitude. (Though towards
the later range the graph appears flat due to lack of resolution)

\begin{figure}[t,h]
\centerline{ \epsfxsize 4in
\epsfbox{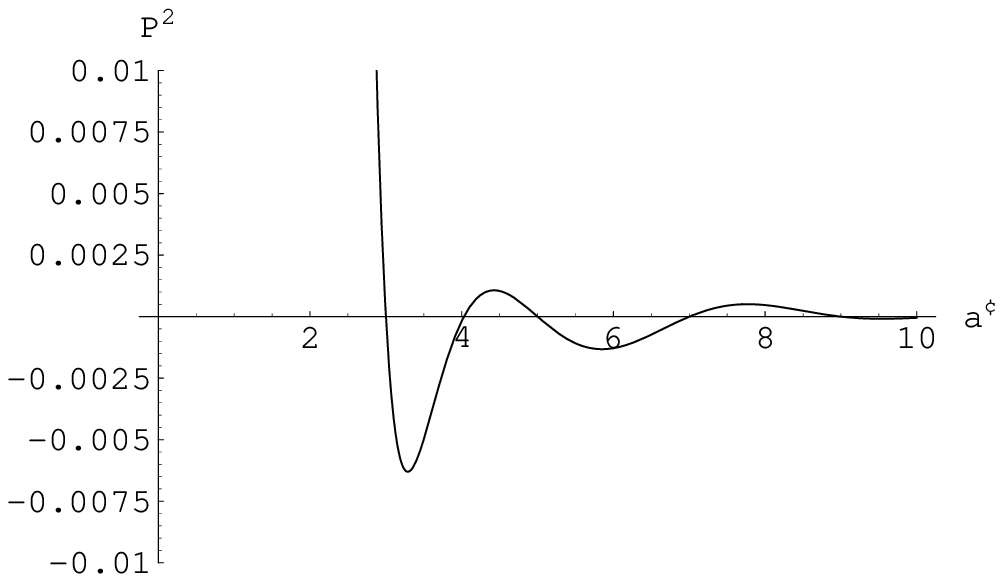}
}
\caption{\sl $P^2_r$: Large size graph}

\label{fig:mom2}
\end{figure}

The plot Figure (\ref{fig:mom2}) is a graph of $P^2_r$ set at $\th_0=\pi/6$ and the size of the dual surface is as large as $\pi/2$. The Oscillations set in as 
quickly as $\a=3$.

A plot of $P^3_r$ is also quite similar to the above. 

The few points to note in the above discussion are:\\
1)The momentum $P_r$ is oscillatory inside the horizon.\\
2)The amplitude is very small, around $10^{-9}$ with $r_g^2/a_N=200$ , just inside the horizon and decreases progressively. Thus to 
prevent the momentum from reaching very low values, one needs to set $a_N = 10^{-9}r_g^2$ at least.
The implications of this are discussed in detail in Section 4.2\\
3)The finer the graph is later the oscillations start, indicating an association with the geodesic deviation
forces which act on the embedded edges.

\subsubsection{Angular edge}

Now we come to the evaluation of the complexified group element for the angular edge.
The angular holonomy is already evaluated along a $\th$ curve in Equation (\ref{holt}). Here we directly evaluate the 
angular edge momentum. Along the plane, it is again a small strip enclosing the radial line along the dual planar graph.

We thus determine $P^I_{\th}$, and this has the form:
\bea
P_{\th}^I &= & -\frac1{a_N}Tr [T^I h_{\th}~\int h_r * E~ h_r^{-1} ~h_{\th}^{-1}] \nn \\
&=& \frac{i}{a_N} Tr[ T^I h_{\th} ~\int \sin{\th} h_r~ r \sigma^2~ h_r^{-1}~  dr h_{\th}^{-1} \nn] \\
&=&\frac{i}{a_N}Tr[T^I h_{\th}~ Y(\th) ~h_{\th}^{-1}] 
\eea

Where when one puts in the expressions for holonomies (\ref{holr},\ref{holt}) , one gets
\be
Y(\th) = Y_2 \sigma^2 + Y_3 \sigma^3
\ee

with 
\bea
Y_2 &= &\sin(\th) \int \cos2\t\left(\frac{1}{\sqrt{r_0}} - \frac{1}{\sqrt r}\right) r dr\\
Y_3 &=& -\sin\th \int \sin2\t \left(\frac{1}{\sqrt{r_0}} - \frac{1}{\sqrt r} \right) r dr
\eea

The above two integrals can be done, and yield 

\bea
\frac{Y_2}{2 \t^4 \sin\th} &=& \frac{1}{24} \left[\cos\left(\a - \frac{1}{\a_+^{1/2}}\right)\left(- \a_+ + 6 \a_+^2\right)
- \cos\left( \a - \frac{1}{\a_-^{1/2}}\right)\left( - \a_- + 6 \a_-^2\right) \nn \right.\\
&-& \sin\left(\a - \frac{1}{\a_+^{1/2}}\right) \left( - \a_+^{1/2} + 2 \a_+^{3/2}\right)
+ \sin\left(\a - \frac{1}{\a_-^{1/2}}\right)\left(- \a_-^{1/2} + 2 \a_-^{3/2}\right) \nn \\
&-& \left. \cos\a\left[ Ci(\a_+^{1/2}) - Ci(\a_-^{1/2})\right] - \sin\a\left[Si(\a_+^{1/2}) - Si(\a_-^{1/2})\right]\right] \label{ang1}\\
\frac{Y_2}{2 \t^4 \sin \th} &=& \frac{1}{24}\left[ \sin\left( \a -\frac{1}{\a_+^{1/2}}\right) \left( - \a_+ + 6 \a_+^2\right) - \sin\left(\a - \frac{1}{\a_-^{1/2}}\right)\left(-\a_ +  + 6 \a_-^2\right) \right. \nn \\&+&  \cos\left(\a - \frac{1}{\a_+^{1/2}}\right)\left(-\a_+^{1/2} + 2 \a_+^{3/2}\right) - \cos\left(\a - \frac{1}{\a_-^{1/2}}\right)\left(-\a_-^{1/2} + 2 \a_-^{3/2}\right) \nn  \\ &-& \left. \sin\a\left[ Ci(\a_+^{1/2}) - Ci(\a_-^{1/2}) \right] + \cos\a \left[ Si(\a_+^{1/2}) - Si(\a_-^{1/2})\right]\right] \label{ang2}
\eea
In the above $\a_+ = 1/\a^2 + \g$ and $\a_- = 1/\a^2 - \g$. ($\g = \d/2r_g$) and $Ci$ and $Si$ denote 
cosine and sine integral functions respectively. 
The momentum is:
\be
P_{\th} = \frac1{a_N}\sqrt{Y_2^2 + Y_3^2}
\ee
The above complicated looking expression can be used to give a equally
complicated expression for the gauge invariant momentum, however when plotted one obtains
a very simple graph for the above.

\begin{figure}[t,h]
\centerline{ \epsfxsize 4in
\epsfbox{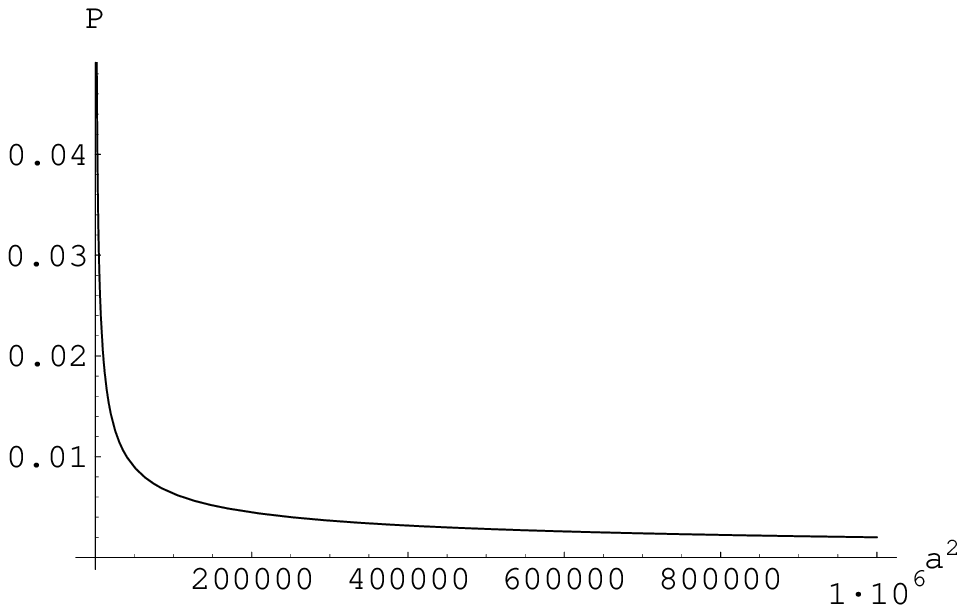}
}
\caption{\sl $P_{e_{\th}}$}

\label{fig:ang}
\end{figure}

The graph in Figure (\ref{fig:ang}) reveals that the behavior of the momentum is damped, but
not drastically as the $P_{r}$, also the singularity does not appear
as any wild point here. This reaffirms the fact that the Momentum, being densitised functions of the dreibein should not be singular at the origin of the
black hole.

Thus we now have all the ingredients necessary to write the Coherent state
wave-function for the planar graph. 

Fine, so the momentum classically can be evaluated, and encodes information about the geodesic deviations which start dominating once
inside the horizon, and inability of $r=const$ surfaces to exist. But what about the coherent state, which is peaked about the above
classical quantities? How do the fluctuations around the above classical quantities depend on the distance from the horizon?
\section{Quantum quantities}
The coherent state as defined
in (\ref{coh}) here, can be written in a series in the irreducible representations. Using the parameterisation for $h= e^{\mu^i T_i}$, where $\mu^I$ denote the angles or three SU(2) parameters, one can write down the series explicitly \cite{thiow1}. As per (\ref{coh}), this involves determining the irreducible representations and the traces corresponding to each for the group element $gh^{-1}$, and then finally summing over all possible representations. However, to see whether the function
converges as $t\rightarrow 0$, i.e. when the width of the state is reduced, one has to use a Poisson re-summation formula.
 This is what was used in \cite{thiow1}, and we quote directly from there. This form, given the complexified element is given as:
\be
\psi_t( g h^{-1}) = \frac{4\sqrt{2\pi} e^{t/8}}{t^{3/2}\sqrt{x ^2 -1}} \sum_{n = -\infty}^{n=\infty} (\cosh^{-1}(x) - 2\pi \imath n) \exp\left\{ - \frac{(2\pi n + i\cosh^{-1}(x))^2}{2 t}\right\}
\ee
Where 
\bea
x &=& \cosh\left(\frac{P}{2}\right) \cos\mu + \imath\frac{P^I\th^I}{P\th}\sinh\left(\frac{P}{2}\right)\sin\mu  \nn \\
 &=& \cosh\left(\frac{P}{2}\right) \cos\mu + \imath\cos(r)\sinh\left(\frac{P}{2}\right)\sin\mu \\
&=& \cosh (s + i\phi) 
\label{def}
\eea
$x$ is the trace of the group element or $x= Tr(gh^{-1})$, and one can choose to write it in a compact form in terms of new variables $s$ and $\phi$ as $\cosh(s + i\phi)$. 

Using the above, one can in principle determine the expectation values of the various operators, defined on this
generic edge. But first, let us study the peakedness properties of this state. Quoting \cite{thiow1}, again we find that
at $t\rightarrow 0$, the so called semi-classical limit, the probability goes as
\be
P^t_H(h) = \left(\frac{16\sqrt{\pi}}{t^{3/2}}\frac{|\cosh^{-1}(x)|^2}{|x^2 -1|} e^{-4(\mu^2 + \d^2)/t}\right)\left(\frac{\sinh P}{P}\right)
\ee
with $\delta^2 = P^2/4 - s^2 + \phi^2 -\mu^2$. Now this is the function of interest in the above probability amplitude. Clearly,
as observed in \cite{thiow1}, this function captures information about the non-abelian nature of the probability.
Thus we concentrate on $\delta^2$ now, as proved in \cite{thiow1}, this function $\delta^2 \geq 0$. Which also implies that as a function of $\delta^2$, the probability amplitude is maximum when 
$\delta^2=0$. Where is $\delta^2=0$?? Amazingly looking at the definition, this is zero when $s = P/2$ and $\phi=\mu$, which
when looking at (\ref{def}) is solved at \be \cos r=\pm 1 ! \label{delta}\ee 

Let us thus look at how the above behaves as a function of the `graph' degrees of freedom.
The behavior of $\cos r$ thus determines the size of the peak, and also the sharpness of
fall around the peak. Let us investigate along the following lines. Due to the fact that
we have taken a Radial edge, and the particular gauge choice of diagonal connections, the
classical holonomy is of the form $h_r = \exp( f(r) T^1)$. Thus when we take the variable
$h= \exp(\mu^I T^I)$, then the $\mu^1$ captures corrections to $f(r)$, while $\mu^2$ and
$\mu^3$ measure corrections along the $T^1$ and $T^2$ directions, where classically the
holonomy component has no contributions. 
Now, let us examine (\ref{delta}), before moving on to other quantities:
\bea
\cos r &= &\frac{P^I \mu^I}{ P \mu} \\
&=& \frac{X_1(r)\mu_1 +\frac{X_3}{\sqrt{\a^2 +1}}\left[ \sin\left(\g\a^3\right) \left(\a\mu_2 +\mu_3\right)+\cos\left(\g\a^3\right)(\mu_2 -\a\mu_3)\right]}{\sqrt{X_1^2 + X_3^2} ~\sqrt{\mu_1^2 + \mu_2^2 + \mu_3^2}} \nn
\eea

\begin{figure}[t,h]
\begin{center}\parbox{3in}{ \epsfxsize 3in
\epsfbox{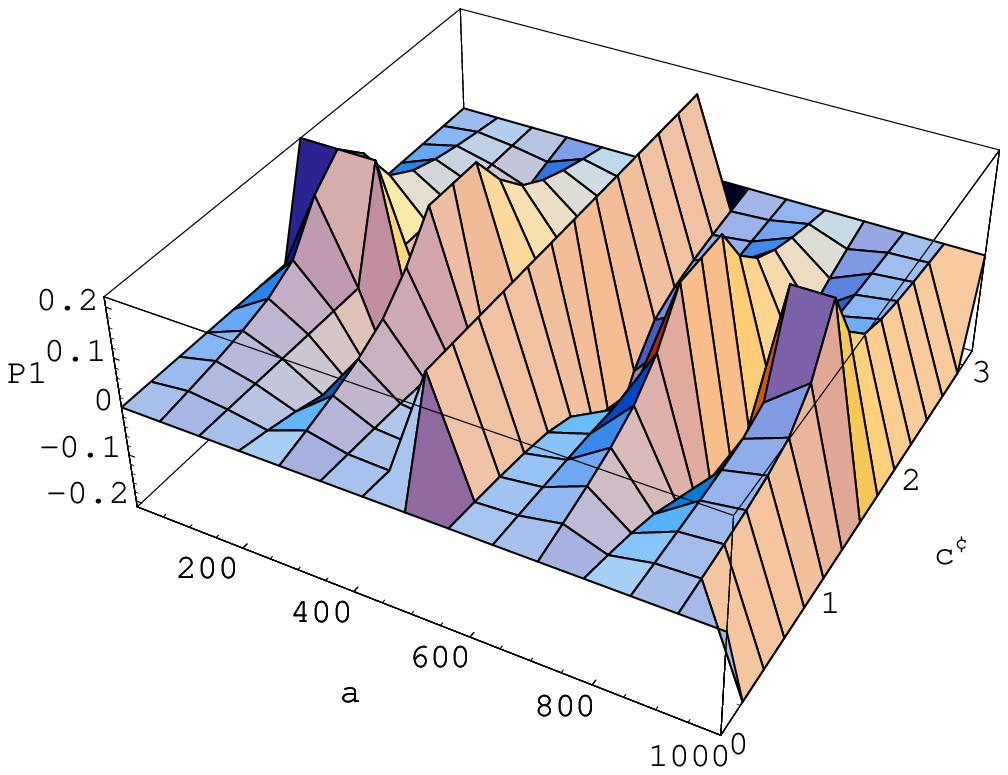}
}
\hspace{.2in}
\parbox{3in}{\epsfxsize 3in \epsfbox{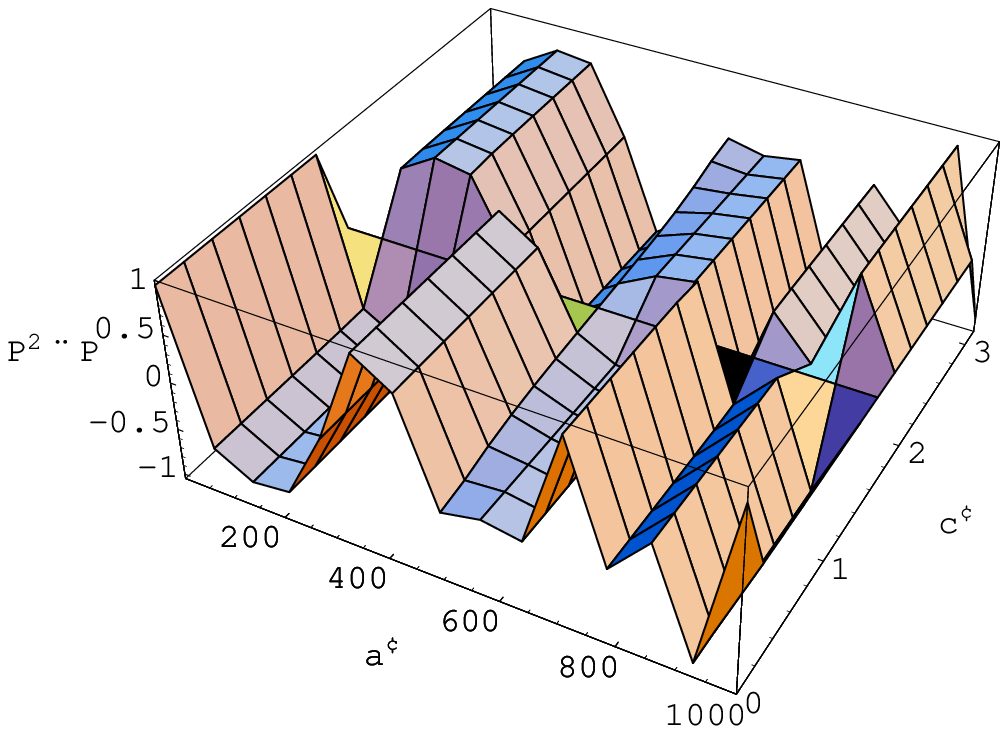}}
\end{center}
\caption{\sl $P^1_r/P_r \ \ P^2_r/P_r$}

\label{fig:quant1}
\end{figure}

First the simple facts: When $X_3=0, X_1\neq 0$, $\cos r = 1$ when $\mu^{1,2}=0$. Thus, when $X_3=0$, the most favored configuration is abelian, i.e. there are no corrections to the classical holonomy along the other two directions even at the quantum stage. However, the existence of such points in the classical phase space {\it are not} independent of the graph degrees of freedom.
e.g. from (\ref{int},\ref{int2}), $X_3=0$ whenever $\th_0= \pi/2$, or the radial edge intersects the dual 
surface at a specified angle! However, this apparently trivial point acquires some meaning when one realizes that one is measuring distances still given by the classical metric, and, it is along
$\th=\pi/2$, one can think that the radial edge acquires the status of a line in flat space: 
$\sin\th=1$, as the 2-surface on which the momentum is evaluated has metric $d\th^2 + d\phi^2$. In so
me sense this is justified as one can think of this as the `ray of least resistance' and as shown in (\ref{momf}), in flat space, the momentum is abelian for the radial edge.
Figure (\ref{fig:quant1})  are the plots for $P^1_r/P_r$ and $P^2_r/P_r$ , while varying the angle at which the 
radial edge is positioned ($\th_0=c'$) and the position of the radial edge itself embodied in
$\alpha$. Now there can be more solutions to $\cos r=1$, and each will come with it's own
physical interpretation in terms of the classical black hole metric. Thus, the coherent state
is very sensitive to classical geometry, and this is precisely what we were
seeking.

Both the plots in Figure(\ref{fig:quant1}) have been evaluated at $\th = \pi/100$, i.e. with a very fine graph indeed,
and confirm our claims with all the points at which $X_1=0,X_3\neq 0$ and $X_1\neq 0, X_3=0$,
appearing at $\th_0= 0,\pi$ and $\th_0= \pi/2$ respectively. However note that in the first
case one does not get $\delta^2=0$.

\begin{figure}[t,h]
\centerline{ \epsfxsize 4in
\epsfbox{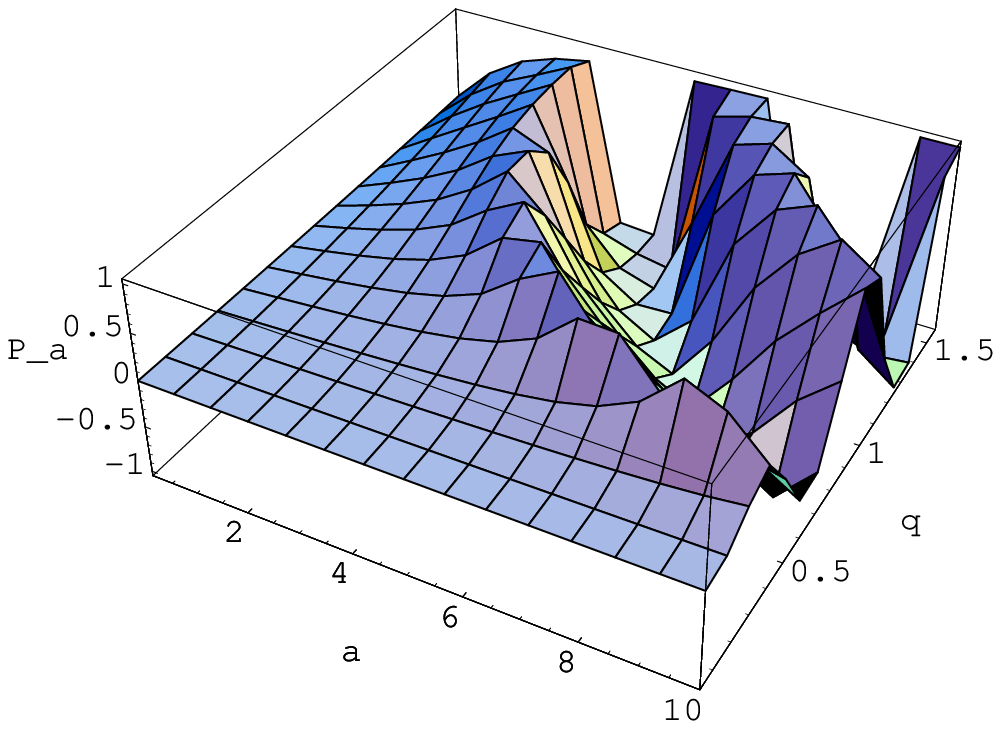}
}
\caption{\sl $P^1_r/P_r$
}
\label{fig:p1}
\end{figure}

Next, we show the plots for as one makes the dual surfaces smaller and smaller with the radial
edge stationed at $\th_0=\pi/6$  Figure(\ref{fig:p1}).
Here, $P^1_r/P_r$ is plotted as a function of the size of the dual surface and also the position at which the dual surface is intersected by the edge, measured as a function of $\a$. Oscillations set in inside 
the horizon, indicating that the solution is no-longer stationary, ($r_g^2/a_N=200$). Note that in all the discussions above, the amplitude is moded out of the P giving finite values for all $\a$.
Thus the dependence in $\cos r$ on $P_r$ is not abnormally damped, and one can study interesting 
behavior for all values of $\a$. The wavefunction is now completely determined, and one has to
evaluate the expectation values of operators in these states, as a function of the distance 
from the horizon. 

\newpage
This is a similar plot for $P^2_r/P_r$, the interesting aspect of this plot was that without the
modulating factor $\sin(\g\a^3)$, Mathematica gave a wild graph at $\a=1$, which prompted
us to think, that we had discovered some new quantum gravity. The graph is attached below
as an example of the many false alarms.

\includegraphics{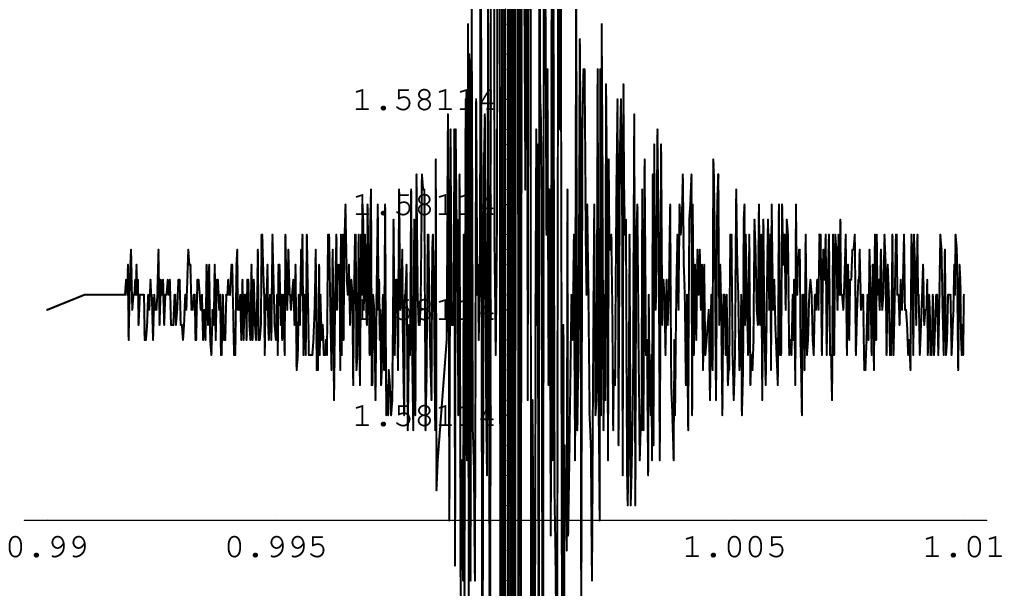}

\newpage

\begin{figure}[t,h]
\begin{center}
\parbox{3in}{\epsfxsize 3in \epsfbox{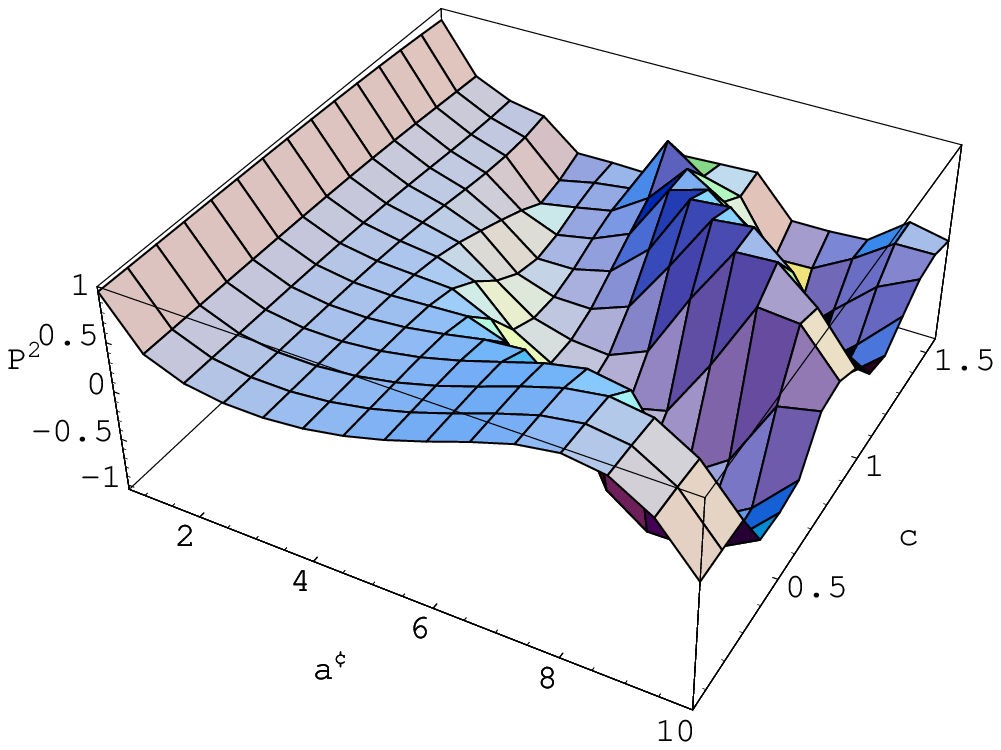}}
\hspace{.2in}
\parbox{3in}{\epsfxsize 3in \epsfbox{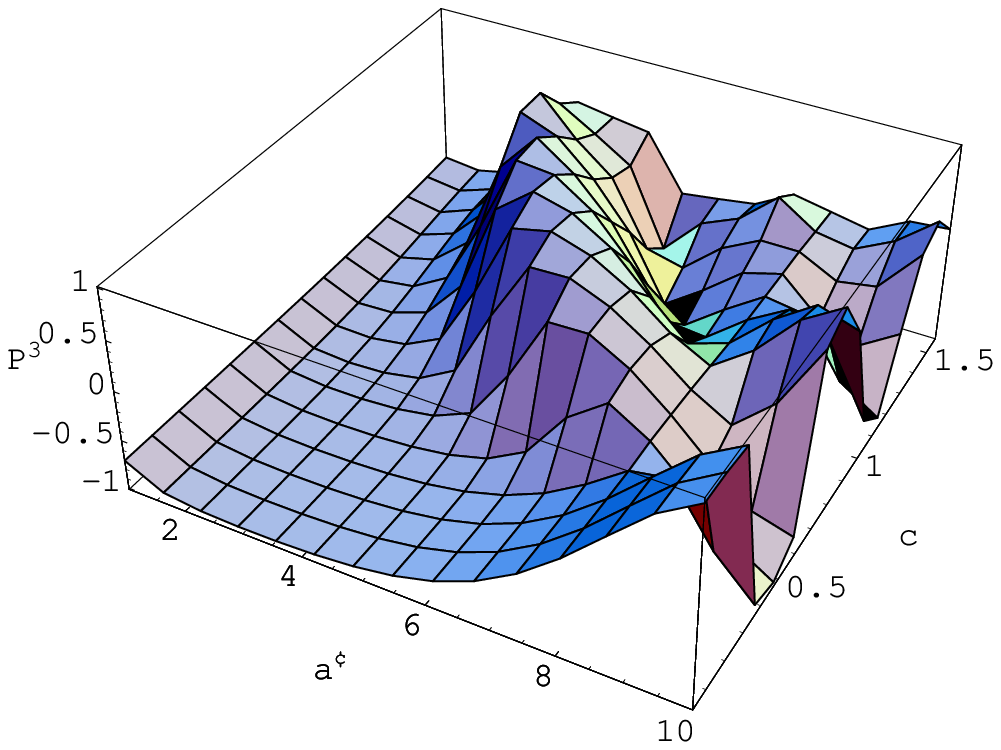}}
\end{center}
\caption{\sl $P^2_r/P_r$ and $P^3_r/P_r$}
\label{fig:p3}
\end{figure}

Adding the proper modulating factor for the $P^2_r/P_r, P^3_r/P_r$ we as expected get good behavior Figure(\ref{fig:p3}). (Note for some reason in Fig. 13 and 14, Mathematica generated $\th$ as $q$ in the .eps file,
and the $P^I_r$'s appear as $P^{(a,b,c)}$ should be read as $P^{(1,2,3)}$.)

All these graphs above, show the existence of certain points at which , the functions become highly oscillatory, and these points
occur {\it always} inside the horizon with their distance decreasing from the centre,
the finer and finer the graph is. Note that making the graph finer allows us to probe shorter
and shorter length scales. Also note that since one has moded the amplitude out,
the functions have values which are quite reasonable. 

Ok, so much so good. But we wanted to calculate corrections to geometry, since anyway we know that the coherent state has the peakedness property as defined above. To do so, we take again the
path of least resistance, i.e. take the radial edge for whom the coherent state behaves as
a Gaussian. 

\begin{figure}[t,h]
\centerline{ \epsfxsize 4in
\epsfbox{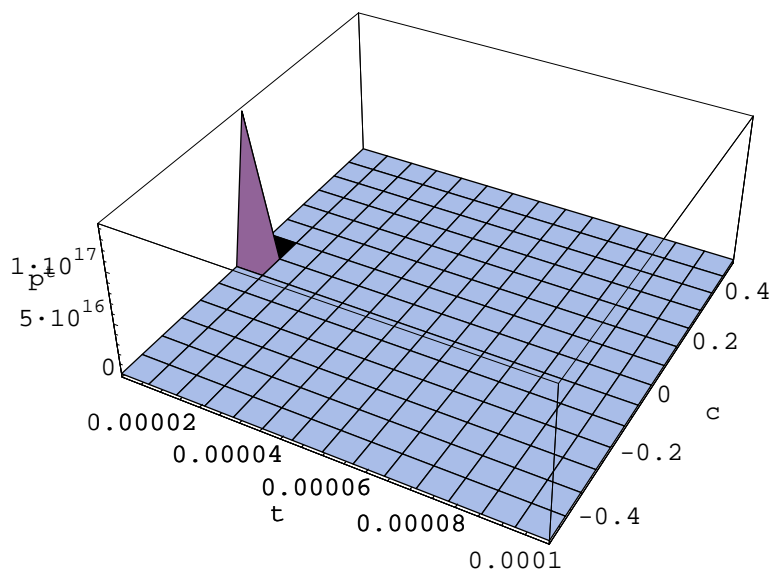}
}
\caption{\sl Probability Amplitude:Peak
}
\label{fig:peak}
\end{figure}

Figure (\ref{fig:peak}) is a plot for $\cos r=1$ and $\delta^2=0$, with the peak appearing as one takes $t\rightarrow 0$, precisely at $\mu=0$

\begin{figure}[t,h]
\centerline{ \epsfxsize 4in
\epsfbox{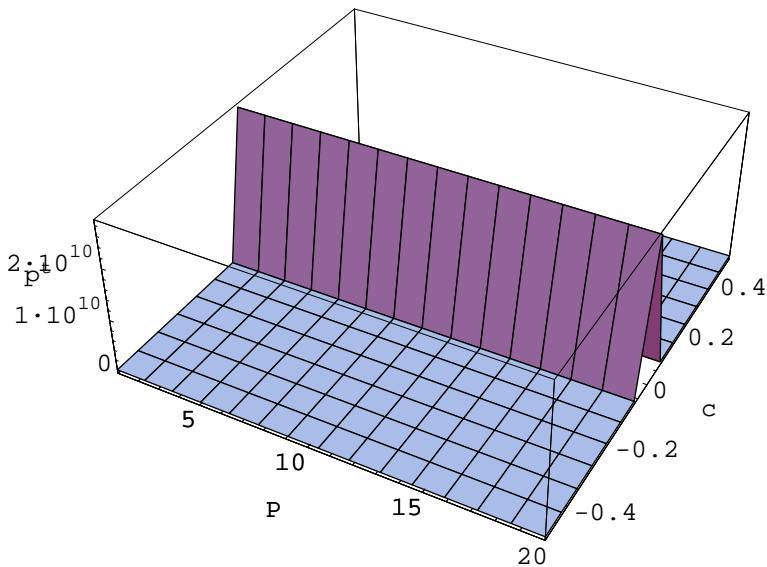}
}
\caption{\sl  Peak Outside Horizon
}
\label{fig:peako}
\end{figure}

Note the probability amplitude is not the probability and hence not restricted below 1. Looking at the probability amplitude, it has the following
form as a function of $P$ and $\m$. 

\be
P^t(h) = \left(\frac{16\sqrt{\pi}}{t^{3/2}}\frac{| P^2/4 + \mu^2|^2}{(\sinh^2(P/2) + \sin^2(\mu))}e^{-4(\mu^2)/t}\right)\left(\frac{\sinh P}{P}\right) \ee
This is almost gaussian in $\mu$, except for the modulating factor in front. To examine that, we plot the above as a function of $(P,\mu)$ Figure(\ref{fig:peako}).
and $t=10^{(-6)}$. The plots show very good behavior.
Next, let us look at $\mu=0$ or the location of the peak in the probability amplitude.
One finds this strange behavior that the peak takes the $P$ dependent value \cite{thiow2}.
\be
\frac{P \sinh(P)}{(\cosh(P) -1)t^{3/2}}
\ee
This is rather surprising, which means that the wave function is peaked higher and higher
as $P$ increases, which is fine for the case of the black hole, where $P$ increases 
polynomially in the radial distance outside the black hole. 
The increase in the probability amplitude implies
that the wavefunction is more `semiclassical' in some sense. 

\vspace{0.5cm}

\begin{figure}[t,h]
\begin{center}
\parbox{3in}{ \epsfxsize 3in
\epsfbox{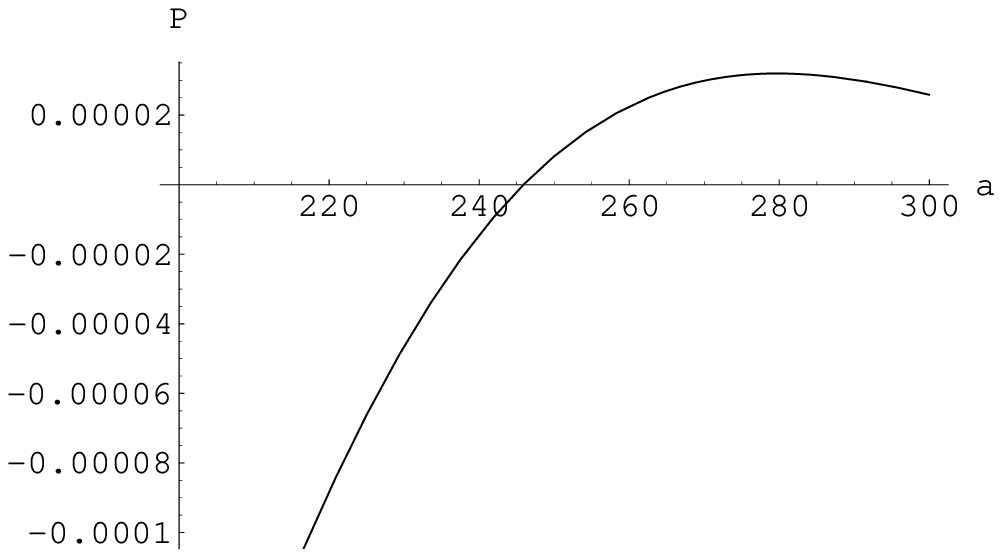}}
\hspace{.2in}
\parbox{3in}{\epsfxsize 3in \epsfbox{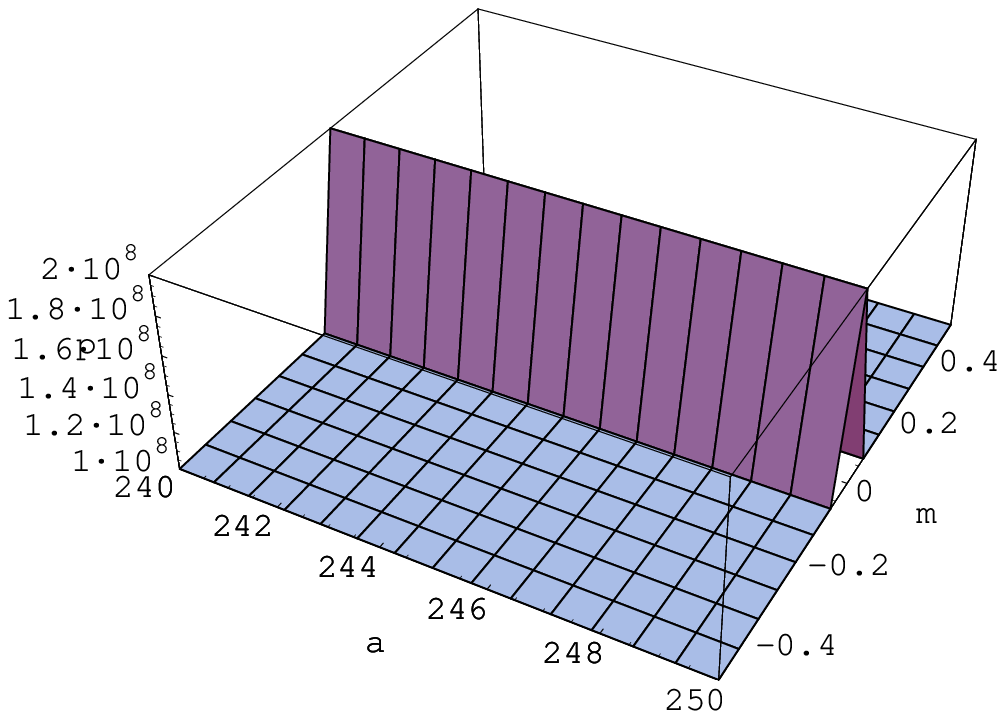}}
\end{center}
\caption{\sl Momentum Inside and Corresponding Peak
}
\label{fig:peam}
\end{figure}

How about the behavior near the point where $P$ oscillates. This is as illustrated in Figure(\ref{fig:peam}).

The first plot is for the $P$, and the next one for the wavefunction around that value of $P$. Note that this
is a plot for black holes with size greater than $10^{9} l_p^2$. Hence as we discuss in section 4.2, the
momentum can be assumed as continuous.

\vspace{.4cm}

So, apparently, everything is fine inside the black hole, along this radial edge. The coherent states appear very robust indeed.

This now brings us to the
calculation of expectation values of the operators. Do the quantum fluctuations
behave nicely? Or there are certain regions in the black hole where they grow to be
very large? The corrections ofcourse shall be proportional to $t$, as computed in
\cite{thiow1}, but they might become uncontrollable in some region, signifying the
breakdown of semiclassical approximations. To determine the corrections, we
see the evaluation of the expectation values of the momentum operator as given in
\cite{thiow2}. The expectation values along the edge of least resistance will have corrections of the form:
\bea
<\psi_g \hat{P}_1\psi_g'> &= & t B  \exp\left[-\frac{(P- P')^2}{4t} -\frac{(\th'-\th)^2}t\right]\left[\frac{\frac{(P+ P')^2}4 \frac{\cosh(P+P')}2 \sqrt{\sinh P \sinh P'}}{\sqrt{PP'}\sinh^2(P + P')/2}\right] \nn \\
&=& t B \exp[-\frac{y^2}{4t} -\frac{\tilde\th^2}t]\left[\frac{(P^2 - P/2y +y^2)(\cosh P \cosh y/2 - \sinh P \sinh y/2)}
{(\sinh P\cosh y/2 - \sinh y/2\cosh P)^2}\right]\nn\\
&\times&
\sqrt{\frac{\sinh P (\sinh P\cosh y-\cosh P \sinh y)}
{ P (P-y)}}
\eea
Where we have defined $y= P'-P$. This is done to extract all the $P$ dependence out such that
one is left with the integral in $y$, which being a dummy variable shall not yield any new
dependence on $P$. This gives for large $P$ corrections
\be
\Delta P = t (P + 1 + 1/P)
\ee

For small $P$

\be
\Delta P = t( P^2 + P +1)
\ee

Thus at least to order $t$, the fluctuations are perfectly under control. Can there be non trivial
results for higher order corrections in $t$? At present we cannot have an answer on it. However,
it is quite a miracle that $P$ is everywhere well behaved, otherwise, existence of such regions
would mean breakdown of the semiclassical approximations. The corrections to the holonomy operator as determined in \cite{thiow2} go as $t$ times a polynomial of the momenta, and hence are finite everywhere. Any operator, be it curvature, are or volume can be expressed in terms of the elementary operators, the holonomy and the momenta. Hence,
quantum fluctuations for them will also be under control.

\subsection{A Quantisation Scheme}
It is interesting to observe that to study the full black hole slice, one 
need to use the Electric field representation. The graph in the configuration space had to exclude the singular point, but the dual graph by construction includes $r=0$. Also, as per equation (\ref{momq}), the singularity at the origin
of the black hole is not seen by the dual momentum as some divergent region.
Hence, to probe this region, one uses the Electric field representation.
The wavefunction in the Electric field representation has been obtained in \cite{thiow1}. 
 Following \cite{thiow3}, the wavefunction in the `electric representation' corresponds to coefficient of one representation $\pi_{j,mn}$ of the Peter-Weyl decomposition. Instead of a continuous valued variable $\mu$ as in the configuration space, one obtains the wavefunction as a function of $j,mn$ corresponding to the Eigenvalues
of the momentum ($j$ corresponds to the eigenvalue of the Casimir $P$, while $m$ and $n$ correspond to the eigenvalues of the left and right invariant $P_3$ \cite{thiow1}). The wavefunction is simply:
\be
\psi_E^t(jmn)= e^{-tj(j+1)/2}\pi_j(g)_{mn}
\ee
The probability appears as:
\be
P^t_E \propto \exp\left\{ -\frac{[(j+ 1/2)t -P]^2}{t}\right\}
\label{propp}
\ee

Where $P$ is the classical gauge invariant momentum, and $j$ gives the eigenvalue
of the Casimir in that particular representation. Clearly for $jt >1$ one can obtain 
a continuous value for $P$. But as mentioned earlier, this is not true when the 
momentum becomes very small indeed. Let us take an example, as evident from
previous section the semiclassicality parameter being $10^{(-9)}$ is enough to obtain the peak.
Whereas inside the horizon, the Momentum oscillates reaching these values very quickly [Figure(\ref{fig:mom34})]. The expectation value of $P$ from Equation (\ref{propp}) is $(j + 1/2)t$, and for one radial edge, one can determine a unique $j$. However, it also means that if there is an adjacent radial edge, with a momentum value
close to it, then $P_r-P_r'= (j_1 - j_2)t$. 
The discrete nature of the spectrum $P=(j+1/2)t$ emerges and only those classical values are `allowed' as peaks in the Coherent State. 

No-matter how big the black hole is, once the oscillations start, the amplitudes decrease very quickly, and soon, one is measuring
values of momenta, which are $O(t)$ even at their zeroeth value, and hence
here, comes the
interesting conclusion, that the Peak values must be discrete!.

\bea
P'_{cl} &= &\frac{r_g^2}{a_N}\frac{1}{\a^4} \left[ \left(\frac{\sin(1-\a')\th}{(1 - \a')}\right)^2 + \left(\frac{\sin(1 + \a')\th}{(1 + \a')}\right)^2 \right. \nn \\
&+& \left.2 \cos 2\th_0 \frac{\sin(1-\a')\th \sin(1 + \a')\th}{(1 - \a^{'2})}\right]^{1/2} = (j + \frac12)t
\eea

Most notable is the presence of the remnant value $1/2$ which prevents the momentum value from ever being zero. This shows that the point $r=0$ where the momentum necessarily vanishes, must be excluded from the
quantum spectrum.
Thus if the black hole is in a coherent state as defined in \cite{thiow2}, then only discrete values of the classical momentum can be sampled by the expectation values. What can we conclude? Quantum mechanics dictates that near the singularity, radial edges can be embedded
only at specific quantised values. Does that mean that space is actually discrete? Will the above quantisation lead to an entropy counting? It is too early to say all that. However it is a certainty that the singularity will be smoothened out, due to the quantum uncertainty principle.

\section{The Actual Graph}
Ok, so much so for the single spoke, but it was needed to set direction to the calculations,
and what quantities to look for, when one considers the actual planar graph. Let us begin by adding more equally spaced spokes, and corresponding dual edges. The actual graph will have tiny radial
edges, tiny spherical edges along $\th$ and $\phi$ directions, and the dual surfaces are corresponding
two surfaces, stretched transversely to the edges. Now, given a vertex of the graph, there will
be three edges starting out along the radial, and the two angular directions and similarly three going in. However, as an act
of simplicity, the $\phi$ angle will again be suppressed. 
So the coherent state at each vertex will be
\be
\Psi_v= \psi_{e^{in}_r}\psi_{e^{in}_{\th}}\psi_{e^{out}_r}\psi_{e^{out}_{\th}}
\ee
As before, the non-trivial information must be in the $\psi_{e_r}$, and we proceed to
extract that. The radial edges are now joined end to end, with dual surfaces intersecting
them at finite distances from each other.

\begin{figure}[t,h]
\centerline{ \epsfxsize 4in
\epsfbox{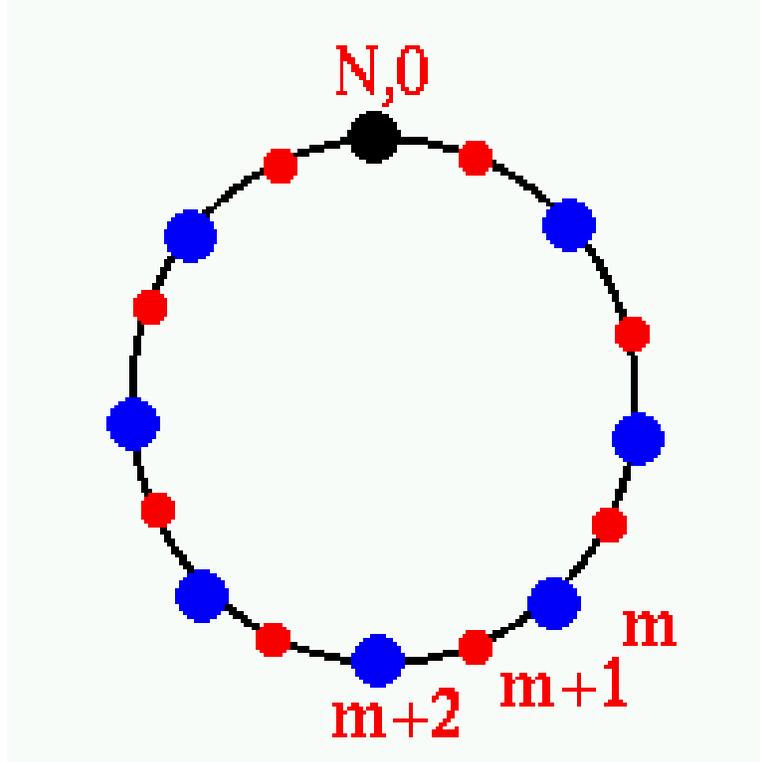}
}
\caption{\sl The Labelled Vertices
}
\label{fig:vert}
\end{figure}

To determine the momenta (\ref{momv}), one has to look at the surfaces, which make up the
sphere now. Choosing a regular lattice/graph, one gets these little dual surfaces 
of equal length for a given radius. The radial edge intersects them at the centre
transversely. Thus on a plane, the cycle consists of N vertices joined by edges of
length $\e/2$ and one unit of dual edge comprising of two such little edges. The
vertex in the midpoint is actually the point of intersection of the radial edge
with the dual edge. This labelling helps in evaluation of the holonomies and the
momenta for the graph.
As illustrated in Figure (\ref{fig:vert}), one starts measuring the distances along
the surface from $\th=0$, and the surface is divided into $N$ bits of 
length $r\e/2$ each, such that $ N r \e/2 = 2 \pi r$. The bits are numbered starting
from $\th=0$ point as $m=0$, and then  $\th +\e/2 $ as $m=1$ and so on. The radial
edge intersects the little surface at $m=2n +1$ odd vertices (denoted in blue dots in the diagram), and the little surfaces
extend from $m = 2n$ to $m=2n +2$ vertices(denoted by red dots in the diagram). 
The value of the radial holonomy is as given in the previous section Equation (\ref{holr}).
For the calculation of the momentum, one needs the holonomies on the surfaces, which are
given by (\ref{holt}), 
where $\th_0$ is the point where the radial edge intersects the two surface, and is denoted by $\th_0= (2n +1)\e$.

The momentum are now given by
\be
P^I_e = -\frac{1}{a_N}Tr\left[T^I h_e \left( X(r)\right) h_e^{-1}\right]
\ee
With $X(r)$ being evaluated for the range of $\th = 2m\e $ to $\th = (2m + 2)\e  $. 
which gives 
\bea
X^m_1(r) &= &\frac{r^2}{4}\left[\sin(2m+1)\e \left\{\frac{\sin (1 -\a')\e}{(1 - \a')} -  \frac{\sin ( 1 + \a')\e}{(1 + \a')}\right\} 
\right]\\
X^m_3(r)&=& \frac{r^2}{4}\left[ \cos(2m+1)\e \left\{\frac{\sin (1-\a')\e}{(1-\a')} + \frac{\sin  (1 + \a')\e}{1 + \a'}\right\}\right]
\eea

Since we have already dealt with the behavior of the above, we donot go into detailed analyses
again. The only thing to take care about is that the radial edges are stationed at discrete
values of $\th$, and the size of the dual surface is fixed at $2\pi/N$, N' denotes the number of the radial edge (N' will be same for radial edges at same distance from the center) . Thus the only variable
is now N given N', and we can choose it to be as large as possible. The momenta will be oscillatory
at greater values of $\a$ as $N$ is increased. Hence finer and finer graphs probe shorter
and shorter length scales.

Also for the angular edge, we again determine $P^I_{\th}$, as determined earlier in Equation (\ref{ang1},\ref{ang2}), however now the limits of integrations
are from $(2n)\delta/2$ to $(2n+2)\delta/2$, with the angular edges intersecting the dual edges at $(2n+1)\delta/2$ radial distances. The dual graph obviously has the zeroeth edge which starts out from the origin. 
Though the fact that the individual components in (\ref{ang1},\ref{ang2}) look singular at the origin, the total expression
for the gauge invariant momentum $P= \sqrt{Y_1^2 + Y_2^2}$ is just given in terms of $\alpha_+$ and $\alpha_-$. Taking $\a_-=0$ and $\a_+ =\g$, one obtains finite terms in the expression given in Appendix 3. ($Lim_{x\rightarrow\infty} \sin (x)/x =0$). Thus, one can write down the Coherent state for the entire planar graph.

Since the individual coherent states are peaked at their respective values, the entire graph will
also be so. The non-trivial expectation values has to come from the evaluation of Volume or Area
operators in the Black Hole Back ground. This calls for a entire new calculation \cite{future}.
Here, we make the following observations. Suppose now, we look at a single radius $r$, and take a
product of the coherent states of all the spokes of the wheel which intersect the radial surface.
This should be a 'spherically' symmetric coherent state. Thus, it would be:
\be
\Psi(r)=\prod_N\psi_{e_{r}}
\ee

As long as one is not evaluating expectation values of operators linking various edges,
one can take the final probability amplitude to be a product of the individual probability
amplitudes:
\be
P^t(r)= \prod_N p^t(h_n)
\ee

And hence one can define an effective probability amplitude, with an effective momentum etc.
The effective momentum goes as $1/{\a}^{4N}$, depending on the number of radial edges one introduces.
Hence, as the graph is made finer and finer, $\a>1$ regions, the value of the momentum decreases
rather fast. The oscillations will now be with a effective size of the sphere which is almost 
macroscopic and set in very near the horizon. In some sense one will be soon in a region of 
microscopic physics for small black holes. The details are under investigation.

\section{Discussion}
In this article, we addressed the coherent state for a black hole. Our main contribution
is the evaluation of the graph degrees of freedom $h_e,P_e$ for a spherically symmetric
graph in a constant time slice of the Schwarzschild black hole. The coherent state by construction is peaked at the classical values of these, and we observed that at certain radial edges,
coinciding with radial geodesics, one gets almost Gaussian behavior of the wavefunction.
Inside the horizon, the geodesic deviation forces become very large, and the $r=const$
spherical surfaces are torn apart, which reflects in oscillatory behavior of the momentum
$P^I_e$. These can be thought of as regions of failure of the graph embedding, and as
one uses finer and finer graphs these regions recede into the black hole. 
Do we get a smoothening out of the black hole singularity? The quantisation condition on the momenta for
small black holes, suggest that indeed space is discrete inside the black hole, in the vicinity of the singularity. 
How robust is the result? How do the graph degrees of freedom affect these, when one tries to average over them? One does not know, but surely, a very strong indication emerges of the physics to come. As of now, we discuss the following points:
\begin{itemize}
\item{The only other `non-perturbative' semi-classical treatment which involves calculations to black hole geometry exists in
2-dimensional CGHS model, though it has matter coupled to it \cite{2dim}. Perhaps, a 2-dimensional version of these
coherent states can be written down, and the results compared with existing literature. A canonical framework for the
2-dimensional model already exists \cite{can2}}.
\item{ The quantisation conditions which arise seem to suggest a degeneracy associated with the black hole, and 
one can get a calculation of Black hole entropy associated with the entire space-time through the coherent state. Will 
it in that case, reveal what the Hawking radiation spectrum should be \cite{spect}? All these and many more issues
involving corrections to entropy, \cite{corr,corr2} are yet to be addressed but the framework now exists.}
\item{The value of $\b$ gets fixed here if one comments on the classical horizon at $r_g=r$, however taking into account
the quantum corrected geometry, one may find $\b$ to be the same value predicted from previous approaches, including the recent quasi-normal mode approach \cite{quasi1,quasi,quasi3}. The details are under investigation \cite{future}.}
\item{Classically, the holonomy and the Momentum variable have oscillatory behavior, as one approaches the singularity.
These are non-local objects, and donot correspond to the usual oscillatory approach to singularity associated with 
scale factors in cosmological models. In ordinary Schwarzschild black hole, the approach to singularity is monotonic
\cite{sch}, however slight perturbations lead to a Oscillatory approach. There is a possibility, that the Momentum
e.g. can be written down as $\int * E'$, where now the $E$ corresponds to a different metric, ($E'$ has all the holonomies
associated with the Schwarzschild absorbed in it), in which case, the oscillatory approach can be associated with
a BKL treatment of singularities \cite{bkl}.}
\item{The original derivation of Hawking radiation involves the Event horizon, which is globally defined. However,
here, the knowledge of few adjacent slices and the presence of the apparent horizon is the only ingredient. Unless
evolution under the Hamiltonian is ensured, one should be careful. BF theory on a single slice has the same results
as Einstein Gravity \cite{hu}, and hence it is important to ensure that inclusion of time evolution does not change the
results drastically.}
\item{There exists semiclassical derivation of the expectation value of the area and volume operator \cite{vol,vol2,vol3} in \cite{thiehanno1,thiehanno2}. The same techniques can be used for the black hole, and corrections to the area operator should be particularly interesting.}
\item{One must couple matter to gravity and repeat the same exercise as most semiclassical derivations of Hawking radiation
and back reaction involve matter fields, and their super-planckian energies in the vicinity of the horizon \cite{back,back1,adg}. Matter
coupled semiclassical states are treated in \cite{thiehanno1,thiehanno2}, and they need to be extended for quantum fields in the curved space-time of
the black hole.}
\end{itemize}
Though the
results reported here need more concrete interpretations, as a begining, there has been many interesting observations, and 
many avenues revealed which need to be explored.

\noindent
{\bf Acknowledgement:}It is a pleasure to thank B. Dittrich, H. Sahlmann and T. Thiemann for initial
collaboration in this project and many lively discussions in the Relativity Tea at AEI, Golm, specially T. Thiemann for useful correspondences and many insightful comments without which the paper would not be written. 
I am also grateful
to J. Barett, S. Das, J. Gegenberg, M. Henneaux, V. Husain, J. Louko, P. Ramadevi, J. Wisniewski for interesting
discussions during various stages of this work, and University of Nottingham for hospitality where this work was
first presented. Thanks to O. Debliquey, B. Knight for help with the computers. 
This Work is supported in part by the ``Actions
de Recherche Concert{'e}es" of the ``Direction de
la Recherche Scientifique - Communaut{'e} Francaise
de Belgique", by a ``P{o}le d'Attraction Interuniversitaire"
(Belgium), by IISN-Belgium (convention 4.4505.86) and
by the European Commission RTN programme HPRN-CT-00131,
in which ADG is associated to K. U. Leuven.  

\noindent
\renewcommand{\theequation}{A-\arabic{equation}}
\setcounter{equation}{0}
\section*{Appendix 1}
The actual graph consists of a foliation of 3-dimensional spheres, and their appropriate duals. The dual
surfaces are 2-dimensional, and involve both the $\phi$ and $\th$ integrals. The radial edge thus
intersects a 2-surface which is part of a sphere sandwiched between two such spherical embeddings of the
original graph. This graph is effectively three dimensional and inclusion of the $\phi$ coordinate adds some complications, but does not add anything drastically new to the observations already made in the previous section.
As usual we begin with a single radial edge, and calculate the corresponding
 three momentum now. The `dual polyhedronal decomposition' of the spherical
geometry now involves pieces of spheres, eg, the dual surface to the
radial edge is piece of a $\th,\phi$ 2-surface. This ensures that the radial
edge intersects it transversely and only once. The projections of the
dual surface onto the plane, gives the edges of the dual planar graph
discussed in the previous section. There is however a subtle ambiguity
associated in choosing the edges of the dual surface which are used to
convolute the triad. One way to fix this is to embed the edges so that
they lie along $\th$ and $\phi$ directions, and then define the path
from any generic point to the point of intersection to be along the
shortest distance curve. However as discussed in the Appendix 2, this
method leads to singular answers for the momentum, and is not suitable
in the study of this problem. One thus resorts to the choice of a path from
a generic point $\th,\phi$ to the point of intersection $\th_0,\phi_0$
to be along a $\phi=\phi$ path which takes $\th\rightarrow \th_0$ and then
moving along $\th=\th_0$ curve such that $\phi\rightarrow \phi_0$. 
The momentum evaluated thus should have now the following integrations:
\be
P^{\rm 3 dim}_{e_{r}} = -\frac{1}{a_N}Tr[T^I h_{e_r}\int h_{\phi} h_{\th} * E h_{\th}^{-1} h_{\phi}^{-1} h_{e_r}^{-1}]  
\ee
Surprisingly $X_2=0$, even in three dimensions, and the origin of that
lies in the choice of the point of intersection to be exactly in the
middle of the dual slice. The $X'_1$ and the $X'_3$ are related to the
2-dimensional ones by the following simple relations:
\bea
X'_1& =& \e X_1\\
X'_3 &= &  \e X_3
\eea
$\e$ is the width in the $\phi$ direction, and it is clear that for $\e$ small the above reduces to (\ref{int},\ref{int2}). The invariant
momentum is different from the 2-dimensional one in it's dependence on $\sin(n\e \a sin\th_0)$,
and has very similar properties. The interesting aspect is that the momentum never depends on the position of the radial edge in the $\phi$ direction, which is a manifestation of the azimuthal symmetry. The momentum for the angular edges is
modified in a similar manner, and one obtaines the coherent state, as a product of the coherent
states for all edges.

\section*{Appendix 2}
In this section, we discuss the ambiguities associated in the choice of edges on the 2-surface
dual to the radial edge. The most appropriate choice of edges would involve the choice of the
shortest path connecting a generic point $\th,\phi$ to the point $\th_0,\phi_0$. This, would
involve if $\th_0<\pi/2$, and $\th<\th_0$, a path along $\th=\th$ such that $\phi\rightarrow
\phi_0$ and then a path along $\phi=\phi_0$ and $\th\rightarrow \th_0$. The situation will be reversed
for points $\th>\th_0$, and one has to travel along $\phi=\phi$ path to $\th\rightarrow \th_0$ and then
along $\th=\th_0$. This ensures that the contribution to the invariant distance comes from $min \sin\th$. Thus, one has to evaluate the integrals of the 2-surface points $\th,\phi$ separately for
$\th<\th_0$ and $\th>\th_0$. Thus for $\th>\th_0$, $h(\th,\phi) = h(\phi) h(\th) = \left[\cos(\frac{\n_0}{2}(\phi_o-\phi))  - i \sigma^3 \sin(\frac{\n_0}{2}(\phi_0 - \phi))\right]\left[\cos(\frac{\a}{2}(\th_0 -\th)) + i\sigma^2\sin(\frac{\a}{2}(\th_0-\th))\right]$ and for $\th<\th_0$, this is 
$h(\th,\phi)=h(\th)h(\phi)= \left[\cos(\frac{\a}{2}(\th_0-\th)) - i \sigma^2\sin(\frac{\a}{2}(\th_0 - \th))\right] \\ \left[\cos(\frac{\n}{2}(\phi_0-\phi)) - i\sigma^3\sin(\frac{\n}{2}(\phi_0-\phi))\right]$, ($\nu_0=\a\sin\th_0$).The Integrals (Note for simplicity all calculations are done for $\a\gg 1$:
\bea
&&\int h(\th,\phi) * E h(\th,\phi)^{-1} \nn \\
& =& r^2 \int(f_0 - i\s^I f_I)\s^1(f_0 + i\s^I f_I)\sin\th d\th d\phi  \\
&=& r^2\int\left[(1- 2f_2^2 - 2 f_3^2)\s^1 + (f_0f_3 +f_2f_1)\sigma^2 - (f_0f_2 -f_1f_3)\sigma^3\right] \sin\th d\th d\phi \nn \\
&=& r^2\int(\cos\a(\th_0-\th)\cos\n(\phi_0-\phi) \sigma^1 + \sin\nu(\phi_0 - \phi)\sigma^2 -\sin\a(\th_0-\th)\cos\nu(\phi_0-\phi)\sigma^3)\sin\th d\th d\phi \ \ \th<\th_0 \nn \\
&=& r^2\int(\cos\a(\th_0-\th)\cos\n_0(\phi_0-\phi) \sigma^1 + \sin\nu_0(\phi_0 - \phi)\cos\a(\th_0-\th)\sigma^2 -\sin\a(\th_0-\th)\sigma^3)\sin\th d\th d\phi \ \ \th>\th_0 \nn 
\eea
The interesting observation is the fact that the integral proportional to $\sigma^2$ is zero for both.
 The integral
proportional to $\sigma^1$ yields the following for $\th<\th_0$ (one takes as the limit: $\phi_0 - n\e/2<\phi< \phi_0 + n\e/2$, $\th_0 - m\e/2 <\th <\th_0 + m\e/2$):
\bea
 && \frac{r^2}{\a}\left[ \sum_{k=0}^{\infty} J_{2k+1} (\a n\e) \left(\cos[(2k +1)\th_0] \frac{2k+1}{(2k+1)^2 -\a^2} \right.\right. \\ &-&\left.\left. \frac{\cos[(2k+1)\th_0 - (2k+1 - \a)m\e]}{ 2k +1 -\a} - \frac{\cos[(2k+1)\th_0 - (2k+1+\a)m\e]}{2k +1 +\a}\right)\right] \nn
\eea
Thus the combined $P_1$ component now has the form:
\bea
P_1 &=& \frac{2 r_g^2}{\a^5}\left[\frac{1}{(2k+1)^2 - \a^2}\left\{ \sum_{k=1}^{\infty} J_{2k+1}(\a n\e) \left[ (2k+1) \cos(2k+1)\th_0 \right.\right.\right.\\&-& (2k+1)\cos[(2k+1)(\th_0 - m\e)]\cos(m\e) \left.\left.\a \sin[(2k+1)(\th_0 - m\e)]\sin(\a m\e)\right\}\right] + \frac{1}{1-\a^2} \nn \\ &\times &\left\{ \left(J_0(\a n\e) -\frac{\sin(n \a \e \sin\th_0)}{\sin\th_0}\right)[\cos\th_0 (1-\cos (m\e)\cos(\a m\e)) - \a \sin\th_0\sin(m\e)\sin(\a m\e)]\right. \nn\\
&+& \left.\left.\left(J_0(\a n \e)+ \frac{sin(\a n\e \sin\th_0)}{\sin\th_0}\right)(\sin\th_0\sin(\m\e)\cos(\a m\e) + \a \cos\th_0\sin(m \e)\sin(\a m\e))\right\}\right] \nn
\eea
However complicated the above expression is, the most important observation is that the expression for momentum
is singular at $\a=2k+1$, and this is not a good sign. The moment one chooses the paths which have symmetric
behavior around the point of intersection, one gets smooth answers. In otherwords, if one chose to consider the choice of path
for $\th<\th_0$ and extend it to $\th>\th_0$ without thinking about the shortest possible way, then one would get for
$P_1$ the following:
\be
P_1 = \frac{r_g^2}{\a^5}\left[\sum_{k=0}^{\infty} J_{2k+1} (\a n\e)\sin(2k+1)\th_0\left[\frac{\sin(2k+1-\a)m\e}{2k+1-\a} + \frac{\sin(2k+1 + \a)m\e}{2k+1 +\a}\right]\right]
\ee
Note in the above, one gets higher modes of oscillations, but the nature is same, the oscillations set in after
$\a=1$ and are highly damped inside the horizon.
Thus, the details of the graph, and edge choices drastically change the nature of the oscillations, what then
is the graph invariant physics?
Obviously, one needs to be careful
about the embedding in the above space-time. However the relief is that, making the graph finer and finer, such that
arguments are smaller $(n\e)$ and $(m\e)$ the ambiguities are reduced even for large values of $\a$, such that
one keeps only the 0th term in the above series.
\section*{Appendix 3}
Here we quote the gauge invariant momentum for the angular direction. We have relegated it to the appendix
due to the length of the expression.
\bea
P_{e_{\th}} &= &\frac{\t^4 \sin\th_0}{a_N}\left[ \left(-\a_+^2 + 6 \a_+^2\right)^2 + \left(-\a_-^2 + 6 \a_-^2\right)^2 + \left(-\a_+^{1/2} + 2\a_+^{3/2}\right)^2 \right.\nn\\
&+& \left(-\a_-^{1/2} + 2 \a_-^{3/2}\right)^2 + \left[ Ci(\a_+^{1/2}) - Ci(\a_-^{1/2})\right]^2 + \left[ Si(\a_+^{1/2}) - Si(\a_-^{1/2})\right]^2 \nn\\
&-& 2 \cos\left(\frac1{\a_-^{1/2}} - \frac1{\a_+^{1/2}}\right) \left\{(-\a_+ + 6 \a_+^2) (- \a_- + 6\a_-^2)
+(-\a_+^{1/2} + 2\a_+^{3/2})(-\a_-^{1/2} + 2 \a_-^{3/2}) \right\} \nn\\ &+& 2\sin(\frac1{\a_-^{1/2}} - \frac1{\a_+^{1/2}})\left\{(-\a_+ + 6\a_+^2)(-\a_-^{1/2} + 2 \a_-^{3/2}) - (-\a_- + 6\a_-^2)(-\a_+^{1/2} + 2 \a_+^{3/2})\right\}\nn\\
&-& 2[ Ci(\a_+^{1/2}) - Ci(\a_-^{1/2})]\left\{ \cos(\frac1{\a_+^{1/2}})(-\a_+ + 6\a_+^2) - \cos(\frac1{\a_-^{1/2}})(-\a_- + 6\a_-^2)\right.\nn\\
&+& \left.(-\a_+^{1/2} + 2\a_+^{3/2})\sin(\frac1{\a_+^{1/2}}) - (-\a_-^{1/2} + 2 \a_-^{3/2}) \sin(\frac1{\a_-^{1/2}})\right\}\nn \\
&-& 2 [Si(\a_+^{1/2}) - Si(\a_-^{1/2})] \left\{\cos(\frac1{\a_+^{1/2}})(-\a_+^{1/2} + 2 \a_+^{3/2}) + \cos(\frac1{\a_-^{1/2}})(-\a_-^{1/2} + 2\a_-^{3/2})\right. \nn\\& +& \left. \left. \sin(\frac1{\a_+^{1/2}}) (-\a_+ + 6 \a_+^2) - \sin(\frac1{\a_-^{1/2}})(-\a_- + 6\a_-^2)\right\}\right]^{1/2}
\eea

\end{document}